 \documentclass[10pt]{article}

\usepackage{graphicx,rotating,hyperref,slashed,amsmath,xcolor,amssymb,amsfonts,colortbl,cite}
\pdfoutput=1
\makeatletter
\hypersetup{colorlinks,bookmarksopen,bookmarksnumbered,
linkcolor=blus,pdfstartview=FitH,urlcolor=rossos,citecolor=verde}
\allowdisplaybreaks

\def\lsim{\mathrel{\rlap{\lower3pt\hbox{\hskip0pt$\sim$}}
   \raise1pt\hbox{$<$}}}         
\def\gsim{\mathrel{\rlap{\lower4pt\hbox{\hskip1pt$\sim$}}
   \raise1pt\hbox{$>$}}}         

 \newcommand{\sfootnote}[1]{}
\definecolor{bluc}{cmyk}{1,1,0,0.1}
\definecolor{rossoCP3}{cmyk}{0,.88,.77,.40}
\definecolor{rosso}{cmyk}{0,1,1,0.4}
\definecolor{rossos}{cmyk}{0,1,1,0.55} 
\definecolor{rossoc}{cmyk}{0,1,1,0.2}
\definecolor{verdes}{cmyk}{0.92,0,0.59,0.4}

\hypersetup{colorlinks, bookmarksopen, bookmarksnumbered,
citecolor=verdes, linkcolor=bluc, pdfstartview=FitH, urlcolor=rossos}

\newcommand{\mio}[1]{}

\definecolor{Gray}{gray}{0.95}

\usepackage{multicol}
\usepackage{color}
\definecolor{rosso}{cmyk}{0,1,1,0.4}
\definecolor{rossos}{cmyk}{0,1,1,0.55}
\definecolor{rossoc}{cmyk}{0,1,1,0.2}
\definecolor{blu}{cmyk}{1,1,0,0.3}
\definecolor{blus}{cmyk}{1,1,0,0.6}
\definecolor{bluc}{cmyk}{1,1,0,0.1}
\definecolor{verde}{cmyk}{0.92,0,0.59,0.25}
\definecolor{verdec}{cmyk}{0.92,0,0.59,0.15}
\definecolor{verdes}{cmyk}{0.92,0,0.59,0.4}

\def\circa#1{\,\raise.3ex\hbox{$#1$\kern-.75em\lower1ex\hbox{$\sim$}}\,}

\newcommand{\nn}{\nonumber}
\newcommand{\beq}{\begin{equation}}
\newcommand{\eeq}{\end{equation}}

\newcommand{\bea}{\begin{align}}
\newcommand{\eea}{\end{align}}
\newcommand{\be}{\begin{equation}}
\newcommand{\ee}{\end{equation}}
\newfam\rsfsfam
\def\mathscr#1{{\fam\rsfsfam\relax#1}}

\def\circa#1{\,\raise.3ex\hbox{$#1$\kern-.75em\lower1ex\hbox{$\sim$}}\,}
\makeatletter

\def\hhref#1{\href{http://arxiv.org/abs/#1}{arXiv:#1}} 

\newcommand{\doi}[1]{\href{http://dx.doi.org/#1}{[doi]}}

\def\hhref#1{\href{http://arxiv.org/abs/#1}{arXiv:#1}} 
 
\def\art{\@ifnextchar[{\eart}{\oart}}
\def\eart[#1]#2#3#4#5#6{{\rm #2}, {\em #3 \bf #4} {\rm (#6) #5} ({\em #1})}

\def\article{\@ifnextchar[{\earticle}{\oarticle}}
\def\oarticle#1#2#3#4#5#6{{\rm #1}, {\em ``#6''}, {\rm #2 #3 (#5) #4}}
\def\earticle[#1]#2#3#4#5#6#7{{\rm #2}, {\em ``#7''}, {\rm #3 #4 (#6) #5}  [\hhref{#1}]}
\def\hepart[#1]#2{{\rm #2, \em#1}}
\def\heparticle[#1]#2#3{#2, {\em ``#3''} [\hhref{#1}]}

\newcounter{alphaequation}[equation]
\def\thealphaequation{\theequation\hbox to
0.6em{\hfil\alph{alphaequation}\hfil}}
\def\eqnsystem#1{
\def\@eqnnum{{\rm (\thealphaequation)}}
\def\@@eqncr{\let\@tempa\relax \ifcase\@eqcnt \def\@tempa{& & &} \or
  \def\@tempa{& &}\or \def\@tempa{&}\fi\@tempa
  \if@eqnsw\@eqnnum\refstepcounter{alphaequation}\fi
\global\@eqnswtrue\global\@eqcnt=0\cr}
\refstepcounter{equation} \let\@currentlabel\theequation \def\@tempb{#1}
\ifx\@tempb\empty\else\label{#1}\fi
\refstepcounter{alphaequation}
\let\@currentlabel\thealphaequation
\global\@eqnswtrue\global\@eqcnt=0 \tabskip\@centering\let\\=\@eqncr
$$\halign to \displaywidth\bgroup \@eqnsel\hskip\@centering
$\displaystyle\tabskip\z@{##}$&\global\@eqcnt\@ne
\hskip2\arraycolsep\hfil${##}$\hfil& \global\@eqcnt\tw@\hskip2\arraycolsep
$\displaystyle\tabskip\z@{##}$\hfil
\tabskip\@centering&\llap{##}\tabskip\z@\cr}
\def\endeqnsystem{\@@eqncr\egroup$$\global\@ignoretrue} \makeatother

\definecolor{fiorentina}{rgb}{.5,0,.5}

\newcommand{\mc}[1]{\mathcal{#1}}

\def\eref{\eqref}

\usepackage[margin=2cm]{geometry}

\begin{document}

\vspace{1truecm}
 
\begin{center}
\boldmath

{\textbf{\Large  Black Holes and Neutron Stars in Vector Galileons}}

\unboldmath

\unboldmath

\bigskip\bigskip

\vspace{0.1truecm}

{\bf Javier Chagoya$^a$, Gustavo Niz$^b$, and Gianmassimo Tasinato$^a$}
 \\[8mm]
{\it $^a$  Department of Physics, Swansea University, Swansea, SA2 8PP, U.K.}\\
{\it $^b$  Departamento de F\'{i}sica, Universidad de Guanajuato - DCI, C.P. 37150, Le\'on, Guanajuato, M\'exico.}\\[1mm]

\date{today}

\vspace{1cm}

\thispagestyle{empty}
{\large\bf\color{blus} Abstract}
\begin{quote}
The direct detection of gravitational waves opens  new perspectives for measuring properties of gravitationally bound compact objects. It is then important
 to investigate  black holes and neutron stars in alternative theories of gravity, since they can have features that  make them
 observationally distinguishable from their General Relativity (GR) counterparts.  In this work, we examine  a special case of vector Galileons, a vector-tensor theory of gravity
with interesting cosmological properties, which consists of a one parameter modification of the Einstein-Maxwell action. Within this theory, we study configurations describing asymptotically flat, spherically symmetric black holes and neutron stars. The set of  black hole solutions in this theory is surprisingly rich, generalising  results found in GR  or in related scalar-tensor theories. We investigate the properties and conserved charges of black holes, using  both analytical and numerical techniques, highlighting configurations that are more compact than in GR. We then  study properties of  neutron stars, showing how the vector profile can influence the star internal structure. Depending on   properties  of matter and fields inside the star, neutron stars can be more massive than in GR, and they can be even more compact than Schwarzschild black holes, making these objects observationally interesting.
We also comment on possible extensions of our configurations to magnetically charged  or rotating configurations.   

\end{quote}
\thispagestyle{empty}
\end{center}

\setcounter{page}{1}
\setcounter{footnote}{0}



\section{Introduction}

The first direct detection of gravity waves opens a new window for astronomy \cite{Abbott:2016blz}, promising to measure with great accuracy properties of gravitationally bound compact objects. In this context, it is important to investigate general properties of black holes and neutron stars in alternative theories to General Relativity (GR).  The aim is to understand whether the physics of compact objects allows one to distinguish among different theories of gravity, which may share similar predictions in the weak gravity regime but present a different behaviour when strong gravity is involved. The simplest and most popular modifications of GR are  scalar-tensor theories, which have been extensively studied in stationary and cosmological situations (see \cite{Clifton:2011jh} for a comprehensive review). Asymptotically flat black holes in scalar-tensor theories are usually similar to those in GR, thanks to  powerful no-hair theorems that forbid the existence of scalar hair in black holes (see  e.g. \cite{Herdeiro:2015waa}). Instead, the physics of compact objects such as neutron stars can reveal interesting new effects, allowing  one to distinguish them from their GR counterparts (see e.g. \cite{Berti:2015itd}).

In contrast, vector-tensor theories are relatively unexplored, although they can have important applications for cosmology. The standard Maxwell electromagnetism, vector field models of inflation \cite{Golovnev:2008cf}, vectors mediating dark forces in  millicharged dark matter scenarios \cite{Holdom:1985ag}, or vector dark energy set-ups \cite{Clifton:2011jh} are examples of this fact. Recently, models of vector Galileons coupled with gravity have been introduced, where the vector  Abelian symmetry is broken by derivative vector self-couplings, and its longitudinal polarization acquires Galilean interactions in the appropriate decoupling limit \cite{Tasinato:2014eka}. These set-ups are free of Ostrogradsky instabilities, and have been explored in a variety of contexts -- see e.g. \cite{Tasinato:2014mia}. In this work, we focus on the physics of spherically symmetric black holes and neutron stars in one of the simplest vector-Galileon scenarios, described by the vector-tensor non-minimal coupling
 \begin{equation}\label{interactionterm}
\beta \,\sqrt{-g}\,G_{\mu\nu}A^\mu A^\nu ,                                                                                 
 \end{equation}
in addition to the standard Einstein-Maxwell Lagrangian. Here, $G_{\mu\nu}$ is the Einstein tensor, $A^\mu$ is the vector field, and $\beta$ is a constant parameter. This term is a special case of these vector Galileons, and the Abelian symmetry is only broken by the non-minimal coupling to gravity. As a consequence, the theory propagates five degrees of freedom: two tensors (gravity), two vectors (the vector transverse modes), and one scalar (the longitudinal vector polarization). The   
latter turns dynamical by the symmetry breaking term \eqref{interactionterm}. We study asymptotically flat black holes and neutron stars,  in the absence of a cosmological constant, pointing out the differences between these systems and their counterparts in scalar-tensor theories, where normally no-hair theorems prevent scalar hair to be detected asymptotically.

The set of black hole solutions in this theory is surprisingly rich, generalising  not only the standard solutions of Einstein-Maxwell theory, but also the black hole configurations of related scalar-tensor theories. When studying neutron stars,  we find that the vector plays an important role for determining the interior configuration of the star, and allows us to find examples of stable neutron stars that can be more massive, and more compact, than their GR counterparts.  Whether the vector field is in the dark sector and plays a role in governing the dark energy or dark matter components, or if it corresponds to a modification of
  Maxwell's electromagnetism in the strong gravity regime,
 our findings may have relevant astrophysical applications.
 
 The roadmap for this work is  the following:
 \begin{itemize}
 \item Section \ref{sec:set} presents our system, some motivations,  and the equations of motion to be examined. 
 \item  Section \ref{blhsec} studies spherically symmetric, asymptotically flat black hole configurations, going well beyond the results we first presented in \cite{Chagoya:2016aar}.  When the parameter $\beta$ is turned on, we find two distinctive branches of asymptotically flat black hole solutions.  The first  branch  is characterised by  non-trivial profiles for the  vector longitudinal and transverse polarizations, and the geometry is well described by a Schwarzschild configuration, plus subleading corrections which decay at large distances. Black holes are characterized by Komar charges corresponding to a mass and a vector charge, similarly to the Reissner-Nordstr\"om configurations in the Einstein-Maxwell theory.  An additional integration constant, mainly controlling the  profile of the  longitudinal vector polarization, is not associated with an asymptotic conserved charge, hence it does not correspond to a black hole scalar hair. In the second branch of solutions,  the configuration is  more 
sensitive 
to the  integration constants associated with vector and scalar profiles; they can modify the leading contributions to the geometrical black hole features, rendering the geometry very different from the Schwarzschild solution. 
 We also study analytically how black hole configurations behave in the limit of $\beta$ going to zero, which requires some care due
   subtle strong coupling effects, which are nevertheless manageable in our context. 

 \item Analytic solutions describing black holes are possible only for special choices of parameters, and more generally, a numerical analysis is needed.  Therefore, in subsections \ref{sec:num} and \ref{sec:bhc} we numerically study some black holes and their properties, in cases where analytic solutions are absent. We analyse what conditions should be satisfied to obtain physically acceptable configurations. 
 We find regions of the parameter space where regular black hole configurations exist, and more interestingly, where the compactness of such solutions can be much larger than in GR, making this vector Galileon model observationally distinguishable from GR in the black hole sector. 
 
 \item  Section \ref{sec:ns} studies neutron stars, with particular attention to the role of the vector profile to specify the star configuration. The non-minimal coupling of equation \eqref{interactionterm} can influence the star's internal structure, since for certain values of the parameter $\beta$, the vector contributes to the energy momentum tensor and modifies the geometry. Moreover, 
  neutron stars in this vector Galileon model can be larger and more massive than in GR, and, for certain parameter ranges, more compact. Actually, they could be even more compact than the Schwarzschild black hole for some cases, making these objects observationally interesting. We also comment on possible generalizations to magnetically charged or rotating configurations, which should be of interest in the case where the vector-tensor coupling \eqref{interactionterm} is considered a modification to Maxwell's electromagnetism, describing exotic objects such as magnetars.

 \item Section \ref{sec:disc} is devoted to conclusions, with a summary of our main results, and a brief discussion of ways forward
 for testing  vector-tensor theories in strong gravity regimes. Moreover, we include appendixes that contain technical details, or extensions to the material
  presented in the main text. 
  \end{itemize}

\section{Set-up}\label{sec:set}

We consider a  specific case of the theory of vector Galileons \cite{Tasinato:2014eka}, which is characterized by the breaking of  an 
Abelian symmetry due to a ghost-free    non-minimal coupling of the vector field to gravity. The action is 
\begin{equation}
\label{eq:action}
S = \int d^4x \sqrt{-g} \left[\frac{1}{2 \kappa} \,R  - \frac{1}{4}F^{\mu\nu}F_{\mu\nu} +  \beta G_{\mu\nu}A^\mu A^\nu + \mathcal{L}_{matter} \right] \,.
\end{equation} 
 $\kappa$ and $\beta$ are the gravitational and vector
Galileon coupling constants,
  $F_{\mu\nu}=\partial_\mu A_\nu - \partial_\nu A_\mu$, and $G_{\mu\nu}$ is
the Einstein tensor.
The dimensions of the quantities involved are 
 as follows:  $R$, $G_{\mu\nu}$  have units of length$^{-2}$, while $\beta$ and $g_{\mu\nu}$ are
 dimensionless.  Using standard QFT conventions, $\kappa$ has units of length squared, and $c=1$.   
 This system is invariant under inversion $A_\mu\to-A_\mu$; however, the  Abelian symmetry, $A_\mu\to A_\mu+\partial_\mu \xi$  (with $\xi$ an arbitrary function) is  broken by the non-minimal coupling between gravity and the vector, weighted by the parameter $\beta$.  
 

 Vector fields  can play an  important role for cosmology,  even hypothetical ones that do not
 correspond to the carriers of   Maxwell
 electromagnetic force. 
The vector-tensor set-up we consider in eq \eqref{eq:action} belongs to a class of theories that have been studied  in various works  as a theory 
of dark energy. Scenarios involving dark vector fields associated with millicharged dark matter and dark photons
 are also sometimes analysed in the context of dark matter. 
 Here we keep agnostic on the nature of the vector field,  and we
 focus on 
 how  it backreacts  to the gravity sector, 
  determining the gravitational properties of compact objects.  Switching on  the non minimal coupling $\beta$ in the action \eqref{eq:action} leads to a far richer variety of asymptotically flat  black holes and neutron stars than configurations associated with the Einstein-Maxwell action.
  


 The covariant field equations of motion (EOMs) corresponding to \eref{eq:action} are
\begin{subequations}\label{eoms}
\begin{align} 
{} 0 &= \frac{1}{2\kappa}  G_{\mu\nu}  - \frac{1}{2}\left[ F_{\mu \rho} F_{\nu}{}^\rho - \frac{1}{4} g_{\mu\nu} F^2 \right]+\beta\left[  
\frac{1}{2}g_{\mu\nu}(D_\alpha A^\alpha)^2 - 2 A_{(\mu}D_{\nu)}D^\alpha A_\alpha  \right. \nn \\
{}  & \quad \left. + g_{\mu\nu} A_\alpha D^\alpha D^\beta A_\beta  +\frac{1}{2} g_{\mu\nu}D_\alpha A_\beta D^\beta A^\alpha   + D_\alpha\left( A_{(\nu}D_{\mu)}A^\alpha +A_{(\mu} D^{\alpha} A_{\nu)}   -A^\alpha D_{(\mu}A_{\nu)} \right)  \right. \nn \\
{} & \quad \left.  - 2 D^\alpha A_{(\mu}D_{\nu)}A_\alpha -\frac{1}{2}\left( A^2 G_{\mu\nu} + A_\mu A_\nu R  - D_\mu D_\nu A^2 + g_{\mu\nu}\square A^2   \right)\right] -\frac{1}{2} T_{\mu\nu} \,,
\label{eeh}
\\
{} 0 & =  D^\mu F_{\mu\nu}  +2 \beta  G_{\mu\nu} A^\mu  \label{veh}\,,
\end{align}
\end{subequations}
where 
\begin{equation}
\label{eq:tmunu}
T_{\mu\nu} = -\frac{2}{\sqrt{-g}}\frac{\delta(\sqrt{-g}\mathcal{L}_{matter})}{\delta g^{\mu\nu}}\, .
\end{equation}
Notice that eqs (\ref{eeh})-(\ref{veh}) do not reduce to Maxwell equations when selecting the Minkowski metric
and $T_{\mu\nu}^{(m)}\,=\,0$.  Dynamical gravity sets constraints that the vector field has to satisfy, even around Minkowski's spacetime	.

In what follows, we are interested in studying configurations that are {\it static, spherically symmetric, and asymptotically flat}. Asymptotic flatness is imposed in order to analyse in a more transparent way large distance properties (such as charges) of the solutions, and to appreciate the differences with scalar-tensor theories.  In the first part of the paper, we focus our study to
 configurations in the absence of a matter energy-momentum tensor (that is, we set ${\cal L}_{matter} \,=\,0 $ in eq \eqref{eq:action}), while in the second part, we study the physics of compact objects corresponding
 to  neutron stars (and we include an ${\cal L}_{matter}$ describing the stellar internal content).

  Under our assumptions,  we select the following general field Ansatz\footnote{One may search for more general static configurations by choosing the possibility of time dependence in the vector field, while keeping the vector's effective energy-momentum tensor and the metric time-independent. However, one of the metric equations (\ref{eeh}) forces the longitudinal component $\pi$ to be a function of  only the radial coordinate, while one of the vector's equations (\ref{veh}) implies that $A_0$ cannot be a function of $t$ and $r$ simultaneously. If time-dependence in $A_0$ is chosen then a further constrain shows that it can only be a constant.} for  determining spherically symmetric configurations
\begin{subequations}\label{ansatz}
\begin{align}
{} ds^2 & = -\left(1 - \frac{2\, n(r)}{r}\right)dt^2 + \left(1 - \frac{2\, m(r)}{r}\right)^{-1}dr^2 + r^2 d\theta^2 +r^2 \sin^2\theta d\varphi^2, \label{eq:metric} \\
{} \textbf A & = (A_0(r), \pi(r),0,0)\,.
\end{align}
\end{subequations}
Notice that we allow for a non trivial profile $\pi(r)$ for the longitudinal polarization of the vector,
which is {\it not} a pure gauge mode in this context, due to the   broken Abelian
symmetry.

\smallskip
Our study aims   to theoretically examine  the properties of regular black holes and neutron stars  in 
scenarios  where the Goldstone boson 
of a broken symmetry has Galileonic interactions, at least in an appropriate decoupling limit \cite{Tasinato:2014eka}. Our vector Galileon set-up shares this feature with the dRGT massive gravity, where black hole solutions have been found (see for example \cite{Tasinato:2013rza}). One advantage of vector Galileons to more complicated theories with vectors such as massive gravity is that black hole configurations are simpler to analyse, and one can analytically understand what happens to these vacuum configurations in the limit where $\beta$, the symmetry breaking parameter, goes to zero. As in other theories where there is no Birkhoff theorem, the black hole geometries found depend on various integration constants, and it is our aim to examine their physical interpretation.

\smallskip
The charged configurations that we will discuss may also have applications in astrophysics. In nature, astrophysical black holes and neutron stars are believed to be {\it uncharged} under Maxwell's electric field, because electrically charged black holes would absorb oppositely charged particles from their surrounding, loosing their own charge. In contrast, if these objects  are   charged under a dark vector field that is different from electromagnetism -- motivated by dark sector
 theories as
 discussed above
--  then the standard model of particles does not feel its associated long range force, preventing a discharge of these configurations.
 \smallskip

\section{Black Holes}\label{blhsec}

In this section, we  determine and analyze solutions corresponding to charged black holes with a regular horizon. To make context with previous work on vector fields in gravity, the
 non-minimal vector-tensor coupling $\beta\,A^\mu A^\nu G_{\mu\nu}$ breaks the Abelian symmetry and acts as an effective mass for the vector on a given background. However, it differs from the usual mass term $m^2A^2$, thus avoiding Bekenstein's no-go theorems \cite{Bekenstein:1971hc} against the existence of black hole solutions in Proca theories. See the discussion in \cite{Chagoya:2016aar} for further details. 
In our set-up, regular solutions do exist for general values of the coupling constant $\beta$ and, both the the vector profile $A_0$ (associated with an electric field $F_{0i}$) and the scalar profile
 $\pi$ (the longitudinal vector polarization), play an important role in characterising these solutions.

When the parameter $\beta$ is set to zero, we recover the Einstein-Maxwell action, where there is a unique static, spherically symmetric, and asymptotically flat solution i.e. the Reissner-Nordstr\"om (RN) black hole configuration
\begin{align}\label{RNsol}
d s^2&=-\left(1-2 M/r+\kappa Q^2/r^2\right)\,d t^2+\frac{d \,r^2}{\left(1-2 M/r+\kappa Q^2/r^2\right)}
+ r^2 d\theta^2 +r^2 \sin^2\theta d\varphi^2
\\
A_0&=\frac{Q}{r}\,,
\end{align}
which corresponds to a charged black hole.
In our case, when the single parameter $\beta$ is turned on, a much richer variety of black hole configurations exist with 
integration constants controlling vector and scalar configurations.  We started the exploration of  this topic in the work \cite{Chagoya:2016aar} by focusing on the specific case $\beta=1/4$. In this section we extend the analysis and show the existence and behaviour of solutions for general values of $\beta$ (see also 
\cite{Minamitsuji:2016ydr,Babichev:2017rti}). 

\smallskip

\smallskip

There are two branches of solutions for our system.  
Substituting our  Ansatz (\ref{ansatz}) in the equations of motion (\ref{eoms}), we find, among other equations, the following
key condition  (obtained  from the off-diagonal components of the Einstein equations):

\begin{equation}
2 \,\beta\,\pi(r)    \left\{n(r)-m(r)-[r-2m(r)] n'(r) \right\}=0\, . \label{constraint}
\end{equation}
When $\beta \neq 0$, 
this equation shows the existence of two  distinct branches of solutions: one with a non trivial profile for the 
vector longitudinal polarization (controlled by the scalar $\pi(r)$); the other
with $\pi(r)=0$. A similar  situation -- two different branches of spherically symmetric solutions -- occur  in many
 examples of modified gravity scenarios, from Gauss-Bonnet gravity \cite{Boulware:1985wk} to 
massive gravity (see for example \cite{Salam:1976as,Tasinato:2013rza,Chamblin:1999tk,Koyama:2011xz}).  An interesting feature of   
 our case is that  both branches admit  asymptotically flat configurations. This implies that there is no uniqueness theorem for spherically symmetric, asymptotically flat 
solutions in this theory. Each branch has distinctive features that we study separately~\footnote{The existence of two branches of solutions has been also pointed out in the recent paper \cite{Babichev:2017rti}, which we received while our work was finalised. (See also \cite{Minamitsuji:2016ydr}.)  Our results are  complementary to those presented in \cite{Babichev:2017rti}, where in addition, we numerically describe properties of black holes in the branch $\pi\neq 0$, analyse properties of vacuum configurations in the branch $\pi=0$, and study neutron star configurations.}.  

Given the special features and physical consequences found in the branch $\pi \neq 0$, we devoted the rest of the main sections to the study of this branch, and leave the discussion of the second branch, where $\pi = 0$, to the appendix \ref{app:pi0}. 
  In Appendix \ref{sec:diff} we instead discuss differences and similarities among vector-galileon and scalar-galileon black holes.

\subsection{Analytical solutions}  \label{sec:analytic}

In the case  $\pi\neq0$, the constraint \eref{constraint} imposes the following relation between the metric components
\begin{equation}
	 m = \frac{n-r n'}{1-2 n'}. \label{eq:mofn}
	\end{equation}
A consequence of this algebraic contraint is that this branch is disconnected from the RN configuration (\ref{RNsol}), which does not satisfy \eref{eq:mofn}.
(RN configurations belong to the second branch of solutions, as we shall see.) Moreover, the polynomial curvature invariants (e.g. the Ricci and Kretschmann  scalars)  vanish at large $r$ for 
arbitrary power-law asymptotic profiles of the metric components when the relation \eref{eq:mofn} holds. This
 {hints towards the  existence of asymptotically flat solutions}.  
	
	In addition, the radial component of Einstein equations is algebraic in $\pi$, with solution 
\begin{align}
	{}\pi^2=&\frac{r}{4\beta(r-2n)}\left[
	2  A_0 (1-4 \beta) \left(\frac{n A_0}{r-2n}+ (r A_0)'\right) -\frac{r A_0^2 \left[1+8 \beta ( n'-1)\right]}{r-2n} 
	-  (r A_0)'{}^2
	\right]\, .\label{eq:pisqr1}  
	\end{align}
The requirement of having a positive right hand side in equation \eqref{eq:pisqr1} will impose constraints on our configurations. 
Together with the Bianchi identity, these conditions reduce  \eref{eeh}-\eref{veh} to two independent
equations for $A_0$ and $n$, which in the absence of matter  reduce to


\begin{subequations}
	\begin{align}
\xi^{(1)}_{vac}
\equiv  &{} \frac{4 \beta-1}{r} \left\{2 r^2 \frac{d}{dr}\left[\frac{A_0 (A_0 r)'}{r}(2 n'- 1)  \right]+2 A_0^2(2 n' -1 )+ A_0 [3 r A_0 -10 r (r A_0)'] n''  \right\} \nonumber \\
	&{}  - \frac{4 (2 n'- 1) n''}{\kappa }-3  [(r A_0)']{}^2 n''+2  (2 n' - 1) (r A_0)' (r A_0)'' =0  \, , \label{eq:00met}  \\
\xi^{(2)}_{vac}
\equiv  &{}   A_0 \left(4 \beta-1\right) n''+ (r A_0)' n''+(1-2 n') (r A_0)''=0 \,  . \label{eq:0vec}
	\end{align}
\end{subequations}
These second order coupled equations admit asymptotically flat black hole solutions, characterized by 
{\it three independent} integration constants: a mass, a vector `electric' charge \footnote{With electric charge we mean the charge associated with the $F_{0i}$ components of the vector's field strength; we do not necessarily identify it with the charge of Maxwell's electric field.} that controls
the profile of $A_0(r)$, and an independent scalar parameter  which controls the profile  of $\pi(r)$. 

Exact solutions to eqs. (\ref{eq:00met})-(\ref{eq:0vec}) describing these black holes are not straightforward  to obtain in general. 
 Nevertheless, one can make progress analytically for some values of $\beta$, or in appropriate perturbative regimes.
There are two special cases where one may find complete analytic solutions in the branch $\pi\neq 0$: one for a specific choice of the coupling constant $\beta$ and the other for a particular value of the vector charge. 

\subsubsection*{Charged black hole solutions for $\beta=1/4$}
For the particular value of $\beta = 1/4$ an exact solution to eqs.~(\ref{eq:00met})-(\ref{eq:0vec}) was found
in   \cite{Chagoya:2016aar}:
\begin{subequations}
\begin{align}
n(r)=m(r)\,&=\,M,  \label{g1f} \\
A_0(r)&=\frac{Q}{r}+P, \label{g1a0} \\
\pi(r)&=\frac{\sqrt{Q^2+2\,P\,Q\,r+2 \,M\,P^2\,r}}{ r-2 M }. \label{g1pi}
\end{align}
\end{subequations} 
The integration constants $M$ and $Q$ represent the mass and electric charge of the configuration. Moreover, $P$ is a new integration constant,  absent in the standard Reissner-No\"ordstrom configuration. It
controls the asymptotic profile of $\pi$, which decays as $r^{-1/2}$ if $P\neq0$, or as $r^{-1}$ if $P=0$. The dimensions of these constants are: $[M]=\rm{length}$, \  $[Q]=\rm{length}/[\sqrt{\kappa}]$ and $[P] = 1/\sqrt{\kappa}$.

This configuration corresponds to the {\it Schwarzschild metric}, but with both profiles for $A_0$ and $\pi$ turned on.
This is an example of a stealth solution, where the effective energy momentum tensor associated to the vector field vanishes and the metric reduces to the GR case. For further details on this solution we refer to \cite{Chagoya:2016aar}. ( See also 
\cite{Minamitsuji:2016ydr,Babichev:2017rti}.)

\bigskip

It is interesting to study how the BH solution departs from the configuration (\ref{g1f})-(\ref{g1pi}), when $\beta$ is  slightly different from $\beta=1/4$. To  understand  this regime, we  build  a series expansion in the parameter $(\beta-1/4)$, with the aim to examine how the location of horizons is affected.
%
  Details on the series expansion can be found in Appendix \ref{app-ch-sec}: here we summarize our findings. The series expansion in $(\beta-1/4)$ is well defined  below some critical radial value $r_{crit}$. As  long as this critical radius is inside the  horizon $r_h$, then we can trust the value of the position of  horizons  calculated perturbatively. The analysis simplifies in regimes where  $P=0$ and $M/Q \ll 1$, where to obtain $r_{crit}\,<\,r_{h}$ one needs
\begin{equation}
   \label{ineco2} 3 \left|\beta-1/4\right|A_0(r_{h})^2 \kappa  \ll 1\,.
   \end{equation}
In this case, the corresponding position of the single black hole horizon is given by
  \begin{equation} \label{posh1}
r_{h} \approx 2 M + (\beta - 1/4)\frac{  \kappa Q^2}{ M } + \mathcal O[(\beta - 1/4)^2]\, . 
\end{equation}
Therefore, as long as our choice of charges and parameters ensures that inequality (\ref{ineco2}) is satisfied at $r=r_h$, the position of the horizon gets indeed shifted when $\beta \neq 1/4$, by an amount depending on the vector charge. Although there is no analytical guarantee that higher order terms in the series do not spoil this result, we performed numerical checks confirming that these findings are correct. 

As a  consequence,  we notice that  vector Galileons can admit black hole solutions which are more compact than Schwarzschild, since they have a smaller horizon radius for the same black hole mass (more on this later). 



\subsubsection*{Solutions describing black holes with no vector charge (but with a scalar profile)}

A second analytic solution in this branch  can be found when the integration constant $Q$ associated with the radial dependence of the vector time-like component vanishes: 
 $Q=0$. The solution is given by
\begin{subequations}
\begin{align}
n(r)=m(r)\,=\,M,  \label{g1f-2} \\
A_0(r)\,=\,P, \label{g1a0-2} \\
\pi(r)=\,
\frac{1}{2} \sqrt{\frac{P^2 (r-2M) + 8 M P^2 r^2}{r (r - 2 M)^2}}.
\label{g1piA}
\end{align}
\end{subequations} 
This configuration is geometrically described
by a Schwarzschild metric, and it is valid for any $\beta$. 
 The solution depends on  two quantities: the black hole mass, $M$, and the scalar parameter $P$. In this uncharged case, this is the same
  solution one finds for a scalar-tensor theory with non-minimal scalar-tensor coupling $\partial^\mu \phi\, \partial^\nu \phi\,G_{\mu\nu}$. We discuss the relation with scalar-tensor 
  theories in full length in Appendix \ref{sec:diff}.


\subsection{Approximate Solutions {and conserved charges}}  \label{sec:appro}

Besides the two analytic solutions we have described, we can  obtain analytic results  by considering perturbative  expansions in
appropriate parameters. We consider two examples with interesting physical consequences. 


\subsubsection*{Asymptotic expansion of solutions for large radial distances, {and asymptotic charges}}

%
%
\noindent
We  analyse the system of equations at large radial distances.
%
   In such regime,  to determine
the solutions for arbitrary $\beta$,  we expand our fields as
\begin{subequations}
\begin{align}  
1-\frac{2 n(r) }{r}\, &=\,1 + \epsilon\, n_1 (r) + \epsilon^2 n_2 (r) + \mathcal{O}(\epsilon^3)\, ,  \label{g1fep} \\
\hspace{2em} A_0(r)\, &=\, P + \epsilon\, a_1 (r) + \epsilon^2 a_2 (r) + \mathcal{O}(\epsilon^3) \, ,  \label{g1a0ep}
\end{align} 
\end{subequations}
where $\epsilon$ is a dimensionless {positive} small parameter,  controlling our asymptotic expansion
in powers of $1/r$. One should stress that the powers can be fractional. The small value of $\epsilon$ reflects 
the fact that we are far away from the source. We find the solutions
\begin{subequations}
\begin{align}
{} 1-\frac{2 n(r) }{r} & = 1-\epsilon \frac{2 M}{r}+\epsilon^2\frac{ Q^2 \kappa \delta  }{2 r^2}+\mathcal{O}(\epsilon^3), \label{gttapp}\\
{} 1-\frac{2 m(r) }{r} & = 1-\epsilon \frac{2 M}{r}+\epsilon^2\frac{ Q^2 \kappa \delta  }{ r^2}+\mathcal{O}(\epsilon^3), \label{grrapp}
\\
{} \hspace{2em} A_0(r) & = P+\epsilon \frac{Q}{r}+\epsilon^2\frac{P Q^2 \kappa \delta  \beta}{r^2} +\mathcal{O}(\epsilon^3), \label{a0app}
\end{align}
\end{subequations}
where the constant parameter $\delta$ is defined by 
\begin{equation}
\delta =(1 -4 \beta )\left({1 }+8 \kappa \beta^2 P^2  -3 \kappa \beta P^2 \right)^{-1}.
\end{equation}
Moreover, 
for the particular case of $P=0 $, the scalar mode has the following form
\begin{equation}
\pi(r) =\epsilon  \frac{Q \sqrt{8 -{1/\beta}}  }{2 r}+\epsilon^2 \frac{M Q (12 \beta -1 )}{2 r^2 \sqrt{\beta  (8 \beta -1 )}}+ \mathcal O(\epsilon^3)  + \dots \label{piappP0},
\end{equation}
while for $P\neq 0 $, it reads
\begin{subequations}
\begin{align}
{} \hspace{2.5em} \pi(r)&  =\sqrt{ \epsilon}\left[\frac{2 P (M P+Q)}{r}\right]^{1/2} \nonumber \\ 
&\quad+\epsilon^{3/2}\frac{\left[32MP(M P + Q)  +Q^2 \left(8 - \beta^{-1} -4 P^2 (1 -4 \beta )  \kappa \delta  \right)\right]}{8 \sqrt{2  r^3 P(MP + Q)}  }+\mathcal{O}(\epsilon^3).\label{piapp}
\end{align}
\end{subequations}
(Recall the dimensions of the quantities: $[M] = [\sqrt{\kappa} Q] = [r]$). As in the exact solution for $\beta=1/4$ of the previous section, $M$ and $Q$ are integration constants representing the mass and vector charge respectively, whereas the parameter $P$ controls the asymptotic profile of the longitudinal vector mode $\pi$. 



\bigskip

This series expansion shows that there are two cases for which  a Schwarzschild metric is 
\emph{exactly} recovered in the weak-field limit: one is for $\beta=1/4$ and the other for $Q=0$, both discussed in Section \ref{sec:analytic} above.
  In all other cases, there are  corrections  which start at second in an expansion in inverse powers of the radius.
 This implies that these corrections are suppressed at large distances with small deviations from GR predictions. As a consequence, the computation of Komar charges related  with the geometry and the conserved vector current 
 $J^\mu\,=\,\nabla^\nu\,F_\nu^{\,\,\mu}$ give the same results as for the standard Reissner-Nordstr\"om configuration 
 (see e.g. \cite{carroll}): the black holes are charaterised by 
  the black hole mass $M$ and the vector `electric' charge $Q$ (with the meaning explained in footnote 3) . The scalar parameter $P$, on the other hand, is {\it not} related
 with a scalar charge associated with any Gauss law that is valid asymptotically. Nevertheless, this parameter is important for determining 
 properties of the solution, as the existence and position of horizons.

\subsubsection*{Small $\beta$ expansion}

In theories where  breaking a symmetry  introduces
  additional degrees of freedom, it is interesting to ask
    whether  there is a well-defined limit where  the symmetry is recovered. The vDVZ discontinuity of the Fierz-Pauli massive gravity \cite{vanDam:1970vg}, or the strong coupling problem in Horava gravity \cite{Charmousis:2009tc} are good examples of how delicate such limit can be, since the theories become strongly coupled. 
 In our scenario,  the equations are sufficiently manageable that a full analytic 
 study can be carried on. 
 A na\"{i}ve $\beta\rightarrow 0$ limit  in the branch of solutions with $\pi\neq 0$ is not  well-defined, because the scalar longitudinal component  $\pi(r)$ of the vector field 
  becomes imaginary. However, 
 a consistent $\beta\to0$  limit {\it can} be defined, if the vector  charge $Q$
  simultaneously goes to zero in an appropriate way.
%
  One should distinguish two cases: $\beta<0$ and $\beta>0$. In the case of $\beta<0$, the fields $A_0$ and $\pi$ admit the following  solution for small $\beta$

\begin{subequations}
\begin{align}
A_0 & = P + \sqrt{|\beta|}\frac{Q}{r} + \mathcal O(|\beta|^{3/2}) \, , \\
\pi & = \frac{1}{2} \sqrt{\frac{Q^2 (r-2M) + 8 M P^2 r^2}{r (r - 2 M)^2}} + \mathcal O(|\beta|^{1/2})\,. \label{eq:pileftlim}
\end{align}
\end{subequations}
This configuration is valid for any $Q$ and $P$ as long as $r > 2M$. On the other hand, if $\beta > 0$ we have
\begin{subequations}
\begin{align}
A_0 & = P + \sqrt{\beta}\frac{Q}{r} + \mathcal O(|\beta|^{3/2}) \, , \\
\pi & = \frac{1}{2} \sqrt{\frac{-Q^2 (r-2M) + 8 M P^2 r^2}{r (r - 2 M)^2}} + \mathcal O(|\beta|^{1/2})\, \label{eq:pirightlim} ,
\end{align}
\end{subequations}
which requires $P\neq 0 $ in order to keep $\pi$ real for any $r> 2M$.
For small $\beta$,
the metric  reduces to a Schwarzschild configuration, plus small  corrections 
\begin{subequations}
\begin{align}
1-\frac{2 n(r)}{r}& = 1-\frac{2 M}{r}+\frac{\beta Q^2 \kappa }{2 r^2}
+ \mathcal O(|\beta|^{3/2}) \, ,
\\
1-\frac{2 m(r)}{r}& = 1-\frac{2 M}{r}+\frac{\beta Q^2 \kappa }{2 r^2}-\frac{\beta \,M\,Q^2 \kappa }{r^3}
+ \mathcal O(|\beta|^{3/2}) \, ,
\end{align}
\end{subequations}

Hence,  we learn that a vanishing $\beta$ limit is connected with a Schwarzschild configuration (and {\it not} with a RN solution), accompanied with a non-trivial
profile for $\pi(r)$, that in any case is non-physical, being a gauge mode in such limit. In Appendix \ref{app:pi0} we  compare this small $\beta$ case with the same
 limit for
  the other
branch of solutions with $\pi=0$.

\subsection{Numerical Solutions }\label{sec:num}

The analytical considerations we made in the previous subsections hint towards the existence of a variety of asymptotically flat charged black holes, 
whose properties depend  on $\beta$, and on the integration constants involved.
 In this section, we numerically explore features of black hole solutions in cases where analytical solutions are not available.
  In Appendix \ref{sec:numext} we analyze the asymptotic behaviour of solutions for large $r$, and even though the analysis is designed
for the neutron star configurations that we will discuss in Section \ref{sec:ns}, the findings remain essentially valid also for the black holes discussed here. One of the main
results of    Appendix \ref{sec:numext} is that asymptotically regular configurations exist only with a relatively small  absolute value  $|\beta|$. For excessively
large values of this parameter, the right-hand-side of eq \eqref{eq:pisqr1} becomes negative at large values of the radial coordinate; hence the scalar $\pi$ turns complex  and the solutions become unphysical. 
  For this reason, in this subsection 
 we choose the representative value $\beta=-1/4$ to  examine  the existence and position of horizons and/or singularities depending on the
 values of  the available integration constants. The value $\beta=-1/4$ is sufficiently different from the $\beta=1/4$ case we have studied at length with analytical methods, and at the same time, is sufficiently small (in absolute value) to avoid scalar singularities  at large radial distances. 

\smallskip

\begin{figure}[ht!]  
  \includegraphics[width=0.45\textwidth]{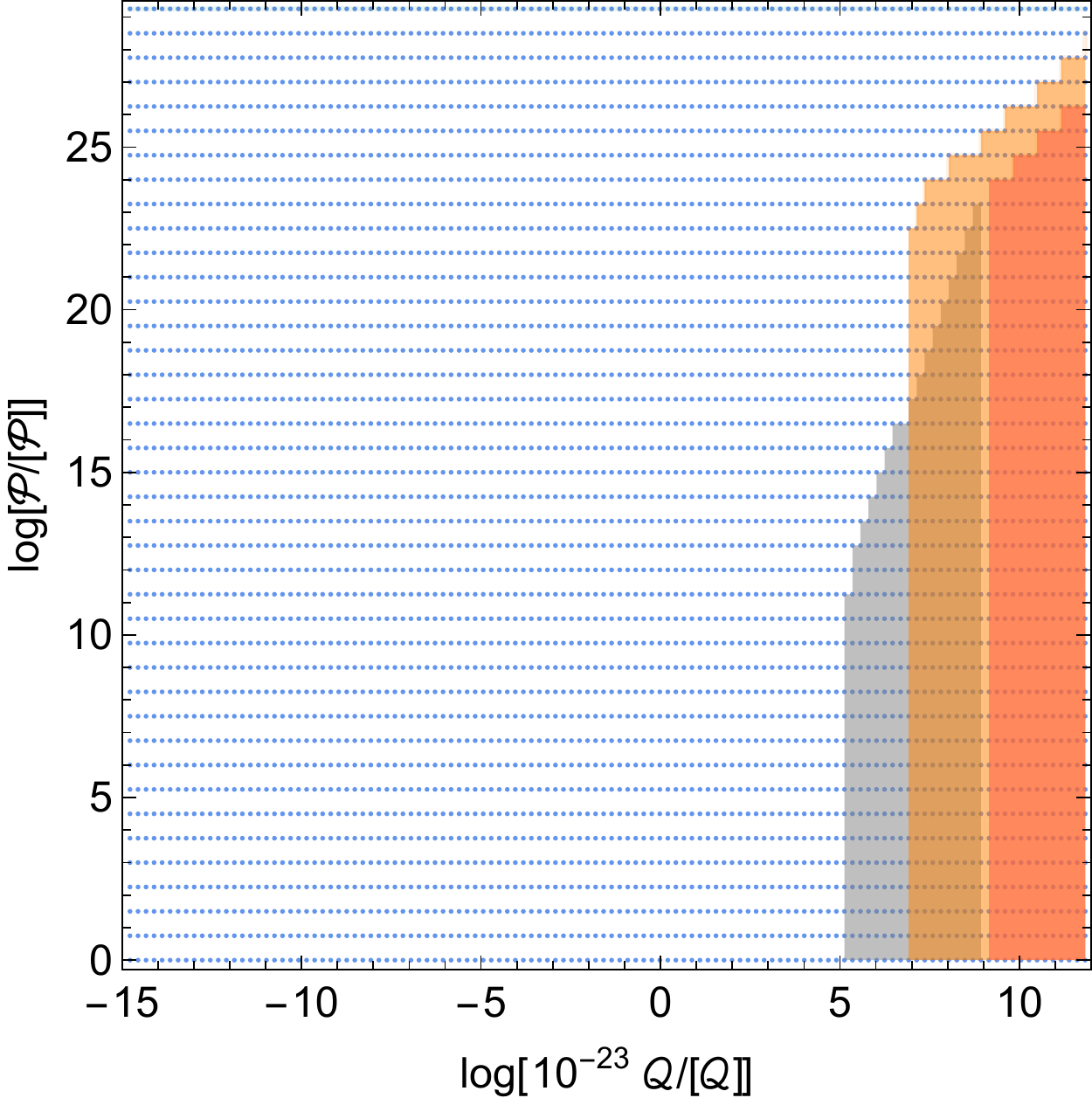} \ \ 	\includegraphics[width=0.455\textwidth]{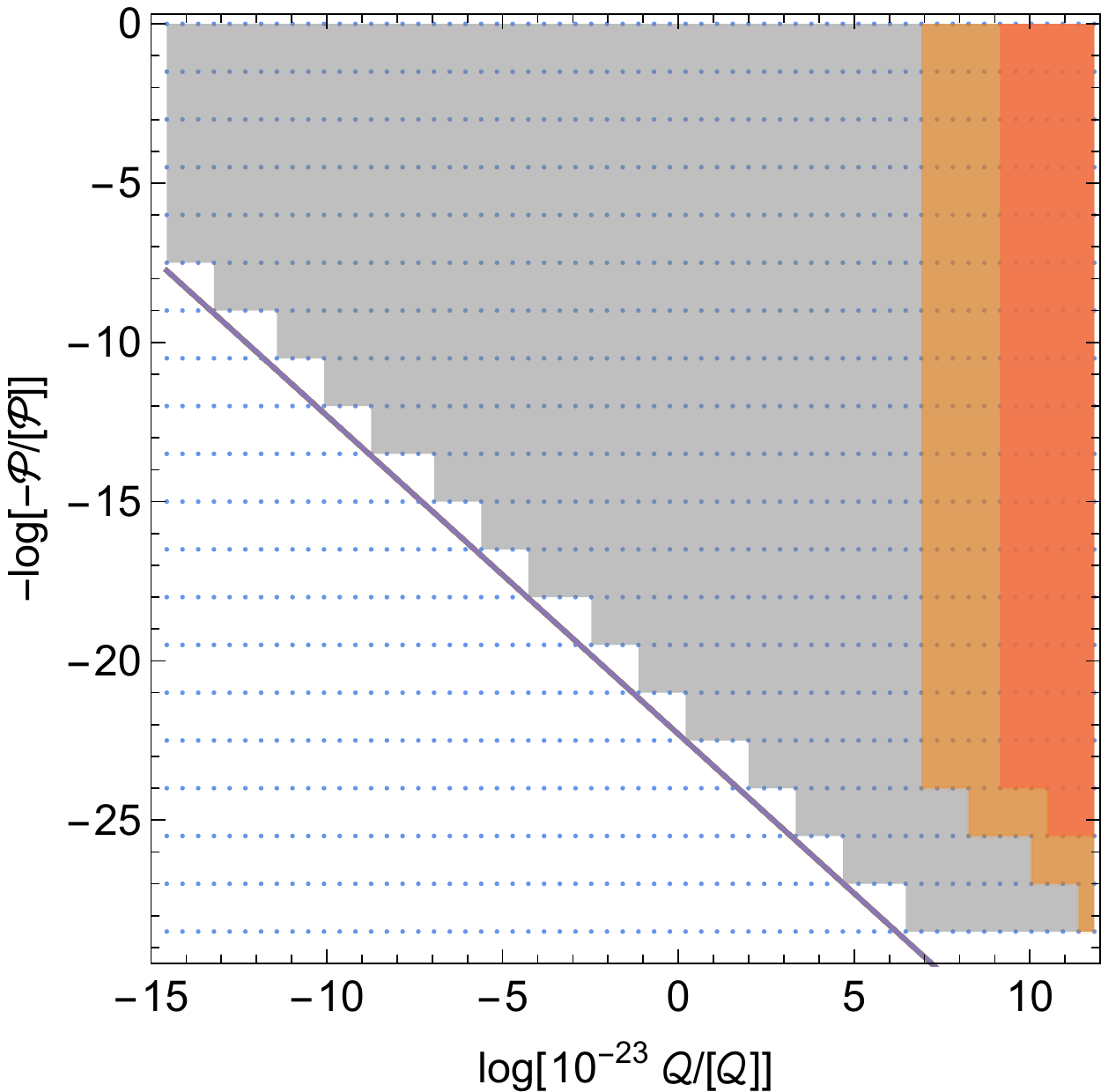} 
\caption{This figure represents the black hole behaviour in terms of the initial conditions $\mc P$ and $\mc Q$ imposed at $r_i = 10^3$~Km. In the horizontal and vertical
axis we use  logarithmic scales: the normalisation factor $10^{-23}$ is introduced for convenience, and the $[\mc P]$ and $[\mc Q]$ factors are used to make the argument of the logarithm dimensionless. Each point in the diagram represents one solution:
 by varying $\mc P$ and $\mc Q$ , we identify four qualitatively different examples of configurations. 
The color notation indicates how the solution behaves: orange (R1) $\rightarrow$ no event horizon, red (R2) $\rightarrow$ singularity at large $r$, grey (R3) $\rightarrow$ complex $\pi$, and blue (R4) $\rightarrow$
regular solutions. 
  All these solutions have $\mc M =4.5 M_\odot$. In the left panel we show the case when $\mc{P}$ and $\mc{Q}$ have the same sign, and in the right panel the case when they have opposite signs. The brown line in the right panel is the analytic limit $MP+Q = 0$ obtained from eq (\ref{piapp}) for solutions with real $\pi$. See the main text for more details. }
  \label{f:inicsA}
 \end{figure}

Hence we study eqs (\ref{eq:00met})-(\ref{eq:0vec}) numerically for  $\beta=-1/4$. We first impose initial conditions on the profile of $g_{tt}$, $A_0$ and their derivative at some  small  value of the radius $r=r_i$. We choose this radius at scale  $r_i \equiv 10^3$~Km, and integrate the equations outwards and  inwards.
  We  do not need to integrate $g_{rr}$, since it is determined by the algebraic constraint (\ref{eq:mofn}). 
 The initial conditions for $g_{tt}$ and $A_0$
can be expressed in terms of three constants, $\mc M$, ${\cal P}$ and ${\cal Q}$, in the following way
\begin{align}
g_{tt}(r_i) &= 1-\frac{2\,\mc M}{r_i} \\
A_0(r_i) &= \mc P + \frac{\mc Q}{r_i},  \\
A_0'(r_i)& = -\frac{\mc Q}{r_i^2}.
\end{align}
Notice that the values
of these integration constants do not necessarily coincide with the black hole Komar charges  at large radial distances
that were discussed in Section \ref{sec:appro}. Indeed, we have seen that
the  metric components and the  time-like vector profile  $A_0$ receive subleading corrections which scale as higher inverse powers of the radius (at most as $1/r^2$). They are sufficient to
modify  the radial values of the  charges. Nevertheless, we numerically checked that the mass $\mc M$ of the configuration is barely affected
by the radial evolution, thus we fix its initial value to $\mc M\,=\,4.5$ solar masses (from now on we denote solar masses with the symbol $M_\odot$), that approximatively corresponds to the asymptotic value of this quantity. To analyze the black hole properties, we vary the quantities ${\mc Q}$, $\mc P$, and explore under which conditions we obtain configurations corresponding to regular black holes.
We look for geometries that are regular everywhere, besides from an essential singularity at the origin, which is  covered by an event horizon located at  the radial coordinate $r_h$ where 
\be
g^{rr}(r_h)\,=\,0\,.
\ee
Moreover, regular configurations require that the scalar field is well defined outside the horizon, without turning complex (as discussed above).
  Our results
are summarised in Figure \ref{f:inicsA}, where we identify four  qualitatively distinct regions of parameters.

\begin{description}
\item[R1 (orange): Asymptotically flat metrics with a naked singularity at the origin:]  The shaded orange region
corresponds to 
  solutions that do not have an event horizon. Their inwards evolution shows that $g^{rr}$ does not vanish at any $r$. Instead the configuration
    diverges as $r\to
  0$, where the Ricci scalar also diverges, indicating that it is an essential
  singularity. For large $r$, the solutions are asymptotically flat.
 
\item[R2 (red): Naked singularities at the origin and at a finite radius:] In the red shaded region we indicate solutions
that  do not have
 an event horizon, and 
  are not well behaved
  asymptotically, since the Ricci scalar also diverges at finite large values of $r$.

 \item[R3 (grey): Black hole solutions with complex $\pi$:] The region indicated in grey includes 
 solutions where $\pi$ becomes complex. 
 We determine them by sampling the value of $\pi$ at
  different distances from the source, for all the solutions with an event
  horizon.

  \item[R4 (blue): Black hole metrics with real $\pi$ everywhere:] This region
   is represented with blue points.  It
   includes all the regular BH solutions with  an event horizon,  and with an everywhere well defined scalar
  $\pi$. 
\end{description}


This numerical analysis shows that the condition of having a regular configuration -- equipped with a horizon and an
everywhere well defined value of the scalar field -- singles out a well defined region in the space of parameters
${\cal P}$, ${\cal Q}$. The boundaries among these  regions in parameter space might contain interesting physics, as
we will see in the next subsection. 

While in this subsection we studied the dependence of the black hole properties from the integration constants ${\mc P}$, ${\mc Q}$ defined at small radial  distances,  we   also repeated the analysis leading to Figure 
 \ref{f:inicsA} considering boundary conditions evaluated far from the origin, where  the corresponding integration constants are more
directly related with black hole charges. We present the details of this discussion in Appendix \ref{app-difpan}.

\subsection{Black hole compactness}\label{sec:bhc}

The black hole compactness, that we refer to by the letter $C$, is an important property to characterise black holes. It is defined in terms of the ratio of the black hole mass versus the position
of the horizon (in Planck units). For the most famous (and relevant for our discussion) black hole solutions its corresponding values are

\begin{align}
 C&=\frac12 \hskip1cm {\text{Schwarzschild BH}}\,,
 \\
\frac12  \le C&\le 1 \hskip1cm {\text{RN BH, upper bound saturated for extremal BH}}
\,.
  \end{align}

  Given that detection of gravitational waves from black hole mergers can allow one to reconstruct geometrical properties of black holes as their size and mass, it is important to examine whether our set-up leads to values of $C$ that can be different from the above ones. 
  
\begin{figure}[ht!] 
  \includegraphics[width=0.65\textwidth]{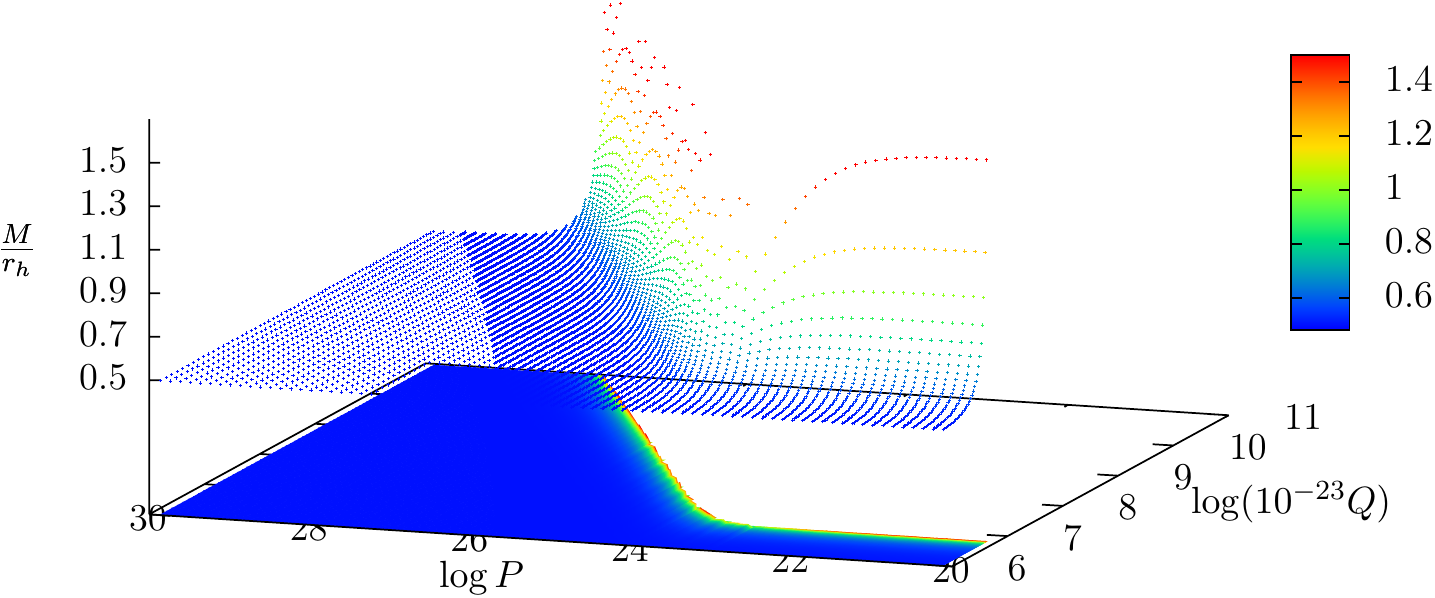}
\caption{Vector Galileon black holes
    with high compactness. For all the solutions in this plot the black hole mass is $M\approx
    4.5 M_{\odot}$. The colour map at the bottom is intended to help to visualise the
    borders of this region and the zones of high compactness, the empty part corresponds
    to {\bf R1} in the notation of Section \ref{sec:num}, where the solutions do not have an event horizon. }  \label{cpq}
\end{figure}

  The answer is affirmative, and  we examine two aspects of
  the question. First, we  focus on the $\beta\,=\,-1/4$ case of Section \ref{sec:num}. 
    If we return our attention to Figure \ref{f:inicsA},  
  at the boundaries between the regular region equipped with horizons,  denoted in blue, and the region where horizon are absent, in orange, we might expect to find values of $r_h$ that are small, suggesting a potentially large $C$.
    This is confirmed in Figure \ref{cpq} where it is shown that by choosing appropriately the values of the available integration constants, we can get configurations with large compactness $C$ --  up to twice the value of RN black holes --  in the proximity of the boundaries between regions.

\begin{figure}[ht!]
	\includegraphics[width=0.46\textwidth]{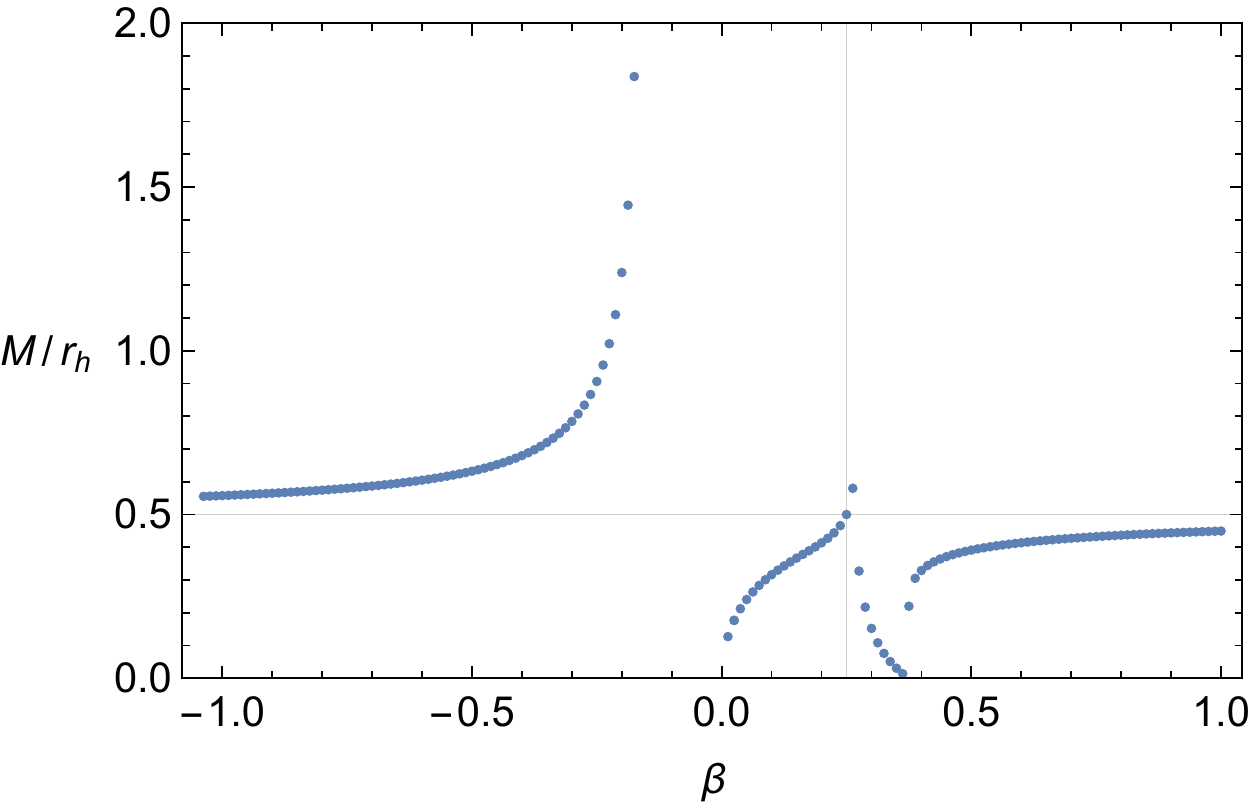} \ 	\includegraphics[width=0.44\textwidth]{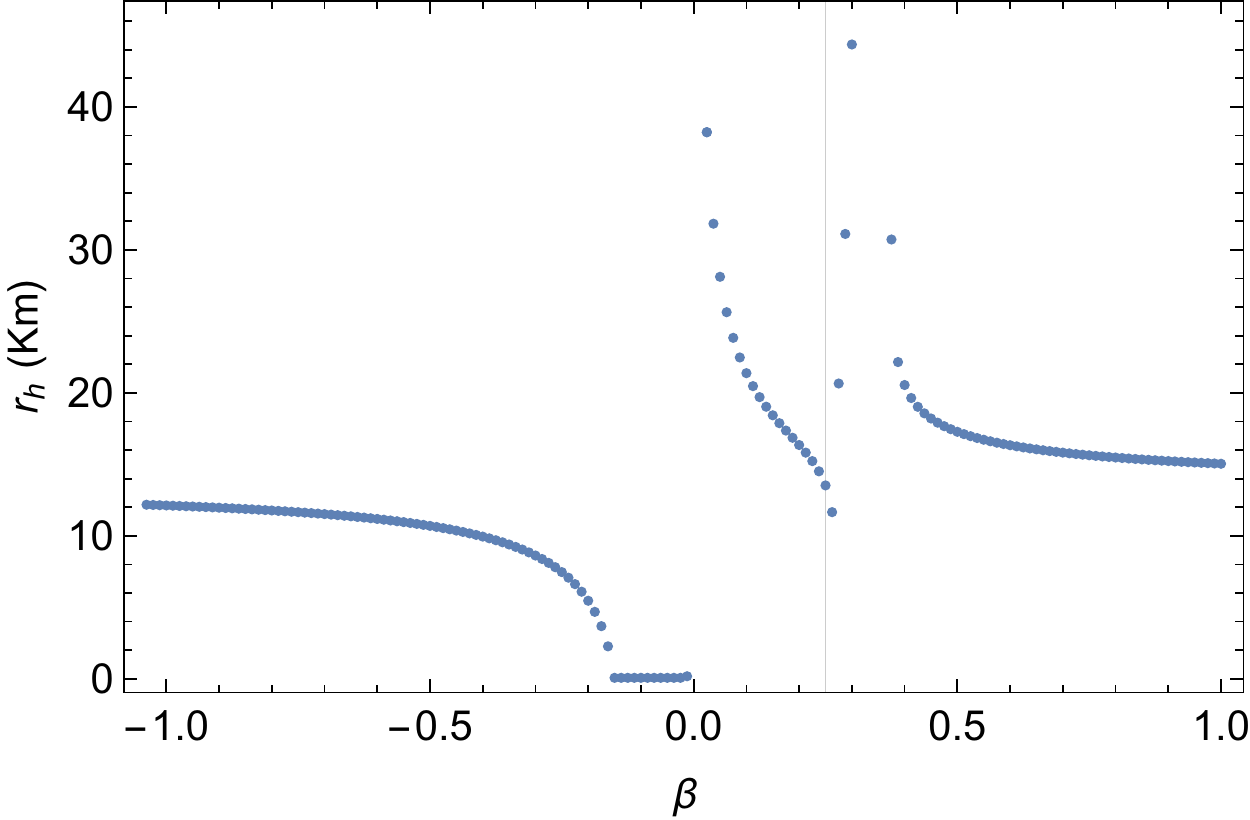} 
	\caption{Compactness for different $\beta$'s at fixed $P=3.1\times10^{17} {\rm cm}^{1/2} {\rm gr}^{1/2}/{\rm s}, Q=4.1\times10^{33}{\rm cm}^{3/2} {\rm gr}^{1/2}/{\rm s}$ and $M=4.5 M_\odot$. The points at $r_h=0$ do not
		correspond to black holes but to naked singularities at the origin. 
	}
	\label{f:cvsbeta}
\end{figure}

  A second possibility  for studying black hole compactness
  consists in  fixing
   the   asymptotic integration constants to representative values and vary the coupling constant $\beta$, while numerically searching for the values of maximum compactness. We present 
   the results of this search  in Figure 
  \ref{f:cvsbeta}, which shows that maximal compactness depends in a non-trivial way on the value of  $\beta$. For
  a choice $\log(10^{-23}Q/[Q]) = 10.6$, $\log(P/[P])=27.5$, and $M=4.5 M_\odot$,   the  compactness is of order $C\,\simeq\, 2$ in proximity of $\beta = -1/4$.

Our numerical analysis shows that 
  that there are regions in the available parameter space where vector Galileons can lead to black holes 
  that are more compact than standard GR configurations.  It would be interesting to further analyse the properties
  of these highly compact black holes, such as their stability and possible observational consequences in dynamical situations, such as binary mergers.


\section{Neutron stars} \label{sec:ns}


Neutron stars (NS) represent a promising arena for studying gravity in a strong field regime, and for 
revealing departures from GR in  modified gravity scenarios (see e.g. \cite{Berti:2015itd,Miller:2014aaa} for  reviews). For example, in the context of scalar-tensor Brans-Dicke theories, Damour and Esposito-Farese have pointed out the phenomenon of spontaneous scalarization
 of NS \cite{Damour:1993hw}. This effect can lead to large deviations from GR predictions, even in theories that satisfy PPN  constraints 
in a weak field regime. The physics of neutron stars have also been studied within the Horndenski scalar-tensor framework.	
 In particular, for a  non-minimal derivative coupling of the scalar field to the Einstein tensor, by tuning the scalar field coupling constants and charge it has been numerically shown that it is possible to obtain NS's with masses up to approximately 2.5~$M_\odot$\cite{Cisterna:2015yla, Maselli:2016gxk}. In this case, the fields outside the NS's are described by the stealth Schwarzschild solution. NS's are also known in Einstein-Dilaton-Gauss-Bonnet gravity (see \cite{Pani:2011xm} and references therein), which is another sub-sector of Horndeski where the scalar field couples linearly to the Gauss-Bonnet term. See \cite{Babichev:2016rlq, Maselli:2016gxk} for  reviews.

In this section we study NS solutions in our vector Galileon scenario, with the main aim to find  NS systems with distinctive properties that make them distinguishable from GR. We show that the additional vector field profile associated with $A_0$ can lead to  consistent NS configurations that are  more massive and more compact than their GR counterparts. We focus on static, electrically charged configurations, and towards the end of this section we also comment on the possibility to include rotation and magnetic fields. This is interesting  in the case where the non-minimal vector-tensor coupling is considered as modification of Maxwell electromagnetism in strong gravity regime.

In order to analyse NS configurations, we need to first determine  solutions for the field profiles in  the exterior of the star. 
These configurations  are then smoothly 
connected with interior solutions,     which depend on the equation of state of the internal matter.
 Interior solutions are determined by  hydrostatic equilibrium configurations  controlled by the matter which forms the star, and by the theory of gravity under consideration -- in our case a vector Galileon model. 
 For definiteness, we focus
on the $\pi\neq 0$ branch of solutions of the constraint condition \eqref{constraint}. In this branch, the geometrical exterior solutions are well described by the Schwarzschild geometry, plus corrections that rapidly decay at large distances, even in presence of non-trivial vector and scalar profiles, as we have discussed in the previous sections. Moreover, the  analysis of Appendix \ref{sec:numext} shows that, for this branch of solutions, the value of $\beta$ can not be too large in absolute value, otherwise the scalar field profile $\pi$ becomes complex in the exterior  of the star.  To avoid this, we make~\footnote{Notice that the analysis of Appendix \ref{sec:numext} uses values of neutron star properties -- radius, mass -- corresponding to neutron star objects similar to GR. As we will learn in this Section, for objects with more exotic  mass or radius one can find unphysical cases where the scalar turns complex even for  $\beta\,=\,-1/4$.}
 the same choice $\beta\,=\,-1/4$ we adopted for the numerical analysis of black hole solutions in Section \ref{sec:num}. For this value of $\beta$
we find  physically interesting NS solutions with sizeable differences from their GR counterparts. We checked
that if $\beta$ is taken positive one does not find configurations that significantly  different from GR.

Before proceeding, 	it is worth to remark that in the main text we do not
consider direct couplings between the vector Galileon field and regular matter. 
In other words, we assume that the only interaction between the star's energy density and the vector field is gravitational.  At the end 
of this section we discuss  options  for  going beyond this approximation. 

	For describing  matter inside a neutron star, we follow Damour and Esposito-Farese work \cite{Damour:1996ke} and use a polytropic Ansatz. We parametrise the equation of
			state in terms of a dimensionless function $\chi(r)$ as 
			\begin{subequations}
				\begin{align}
				\rho(r) &= \rho_0\left( \chi +\frac{ K}{\Gamma -1} {\chi^\Gamma} \right),\label{eosrho} \\
				p(r) &= K\rho _0 \chi^{\Gamma},\label{eosp}
				\end{align}
			\end{subequations}
			with
			$\rho_0 = c^2 n_0 m_b=c^2 1.66\times 10^{14}{\rm gr}/{\rm
				cm}^3$.  $n_0$ is the baryon number density and $m_b$ the baryon
			mass. The
			  polytropic constant $K$ and exponent $\Gamma$ take values appropriate to
			adjust the observed masses of NS. In GR, the values{ }\footnote{We re-adjust the value $K=0.0195$ used in \cite{Damour:1996ke} in
				order to account for updated observations on the maximum mass of neutron stars.} $\Gamma=2.34$ and
			$K=0.0225$ give a mass-radius (M-R) curve with a maximum
			gravitational mass of {2.05} solar masses, in agreement with
			the largest observed neutron star masses (PSR\,J1614-2230 with  $M=1.97\pm0.04$ Solar masses  
			\cite{Demorest:2010bx} and PSR\,J0348+0432 with  $M=2.01\pm0.04$
			Solar masses \cite{Antoniadis:2013pzd}).   We have verified that the qualitative features of our configurations do no change
			  if a more accurate equation of state  is considered. 
 
 The covariant equations of motion (\ref{eoms}) in the case of the neutron star configurations we are considering simplify in the following way. In the interior of the star, there are two algebraic conditions which determine the profiles for the metric component $m(r)$ and the scalar field $\pi(r)$, given by
 	\begin{subequations}
			\begin{align}
                          m(r) =& \frac{n-r n'}{1-2 n'}\, , \label{metricconstraintp}\\
                          \pi^2(r)=&\frac{r}{4\beta(r-2n)}\left[2 r^2 (1-2 n')p
                            - 2  A_0 (4 \beta - 1 ) \left(\frac{n A_0}{r-2n}+ (r A_0)'\right)\right. \nonumber \\
                          & \left.\hspace{0em} -\frac{r A_0^2 \left[1+8 \beta (
                                n'-1)\right]}{r-2n} - (r A_0)'{}^2 \right]\, .
                          \label{eq:pisqrp}
			\end{align}
                      \end{subequations}
Moreover, we have dynamical equations for the remaining quantities, which have the same structure as in the exterior of the object, but with contributions associated with the star energy momentum tensor, namely
\begin{subequations}
			\begin{align}
                          {} 0 = &  \xi^{(1)}_{vac}   -\frac{2 r^2 \rho (2 n'-1)^3}{r-2 n}  +2 p r (2 n'-1) \left[\frac{(2 n'-1) \left(4 n-3 r+2 r n'\right)}{2 n-r}-r n''\right]  \, , \label{eq:00metlp} \\
                          {} 0 = & \xi^{(2)}_{vac} \, \label{eq:0veclp} .
			\end{align}
			\end{subequations} 
			where $ \xi^{(1)}_{vac}$ and  $\xi^{(2)}_{vac}$ are defined in eq \eqref{eq:00met} and \eqref{eq:0vec}.		 Together with the equation of state (\ref{eosrho}-\ref{eosp}), and the energy-momentum tensor
			conservation, $\nabla_\mu\,T^\mu_{\,\,\nu}\,=\,0$, 
			 these equations further determine the time-like component $A_0$ of the vector,
                         the $00$-component of the metric (the function $n(r)$), and the density and
                        pressure of matter. 
                         
\begin{figure} 
                         \includegraphics[width=0.85\textwidth]{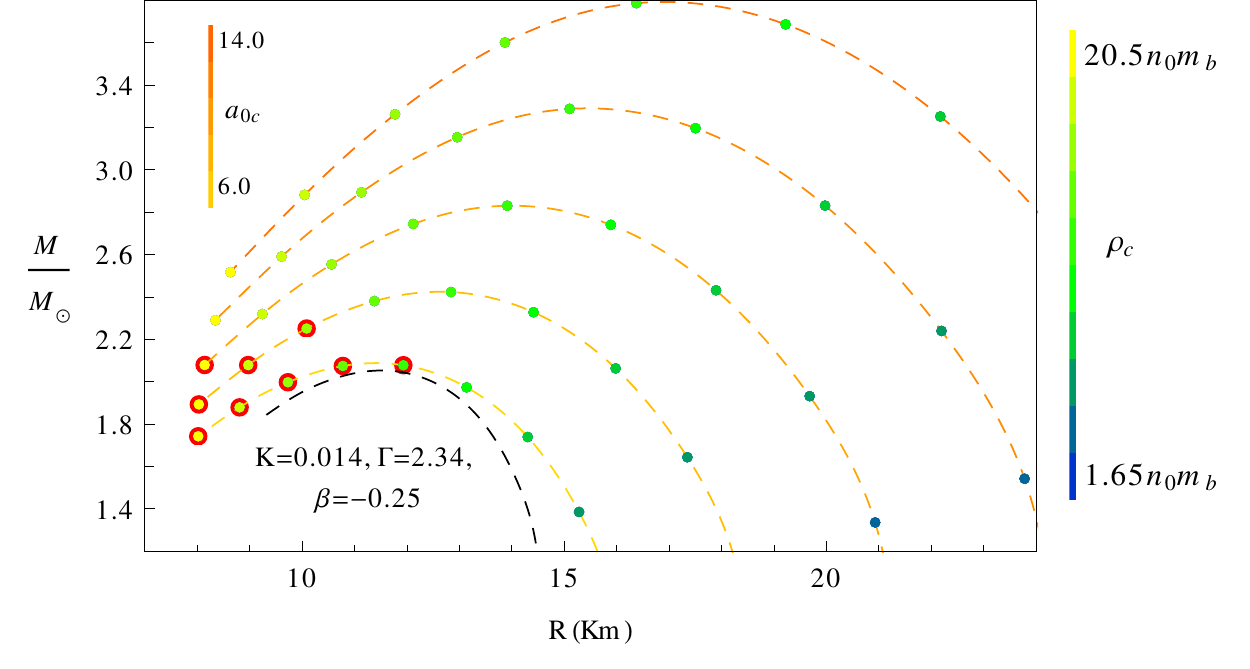}
			\caption{Mass-radius curves for neutron stars in vector Galileons for $\beta=-1/4$. We use a polytropic EoS with $K=0.014$ and
			$\Gamma = 2.34$ . Each line
                            corresponds to solutions with the same central value $a_{0c}$
                            indicated by the colour bar on the top left corner of the plot,
                            and each point along these lines corresponds to a solution
                            with central density $\rho_c$ indicated by the sidebar.  The dark dashed line is the $M$-$R$ curve adjusted in GR to fit the 
                            maximum observed mass of a NS, it is computed using a polytropic EoS with
                            	$K = 0.0225$ and $\Gamma = 2.34$. This
                            plot shows that for  $\beta=-1/4$ we can obtain
                            equilibrium configurations with values of $m(R_*)$ larger
                            than in GR.  Circled points correspond to unphysical configurations where the scalar field
                            turns complex inside the 
                          star radius.}
				\label{ns-vgbetaminus1}
			\end{figure}

                         We numerically solve these equations imposing appropriate boundary conditions, 
                          and choosing  the value $\beta\,=\,-1/4$ for the  vector Galileon coupling constant. The condition
                         that the metric is well behaved at the center of the star, $n(r=0)=
                        0$, requires $A_0'(0) = 0$. We are left with only one free parameter 
                          to fully determine  the initial condition at the origin: 
                          $ A_0(r=0)\,\equiv\, A_{0c}$.  
 To work with dimensionless quantities,  it is convenient to 
                        parametrise the initial conditions for $A_{0c}$ as $$a_{0c}\,=\,A_{0c}/  (10^{23} {\rm cm}^{1/2}{\rm gr}^{1/2}/{\rm s})\,, $$ so that $a_{0c}$ is  dimensionless and its logarithm of order 1.                     
                        Our aim is to determine how NS configurations depend on $a_{0c}$, and how the results differ from
                        standard GR configurations. We match the interior solution at a star radius $R_\star$ with a regular, asymptotically flat exterior solution for the field equations. 
                        The radius $R_\star$ corresponds to the value of the radial coordinate where the internal pressure vanishes. 
                        Figure   \ref{ns-vgbetaminus1} shows 
                        our results in terms of mass-radius (M-R) curves for
different initial conditions $a_{0c}$. Vector Galileons allow neutron star masses and
                          radii larger than in GR,
                          whose permitted configurations are represented with a black 
                          dashed line in Figure   \ref{ns-vgbetaminus1}. 
                          The field profiles are well defined for any $a_{0c}\ge 5$, and our configurations are stable according to the
                          static stability criterion $dM/d\rho_c>0$
                          \cite{Haensel:2007yy}. 
                          For $a_0\gtrsim 10$ we get
                            masses larger than the upper bounds theoretically estimated in
                            GR for any EoS \cite{Rhoades:1974fn}. Given the additional degree of freedom, it is expected that modified gravity models may help to relax the GR mass bounds 
                            \cite{will1993theory}.


	\subsection{Neutron star configurations when varying 
	 $a_{0c}$}		
			   
			   The initial condition corresponding to the value of the 
			   vector profile at the origin, 
			     $a_{0c}$,  represents  the key 
			   parameter for controlling the NS solutions. As we
			   decrease $a_{0c}$, maintaining the same values for the EoS parameters, we find that the
			   scalar $\pi$ turns complex, hence the configuration has to be discarded. 
			   %
%
                                
\begin{figure}[ht] 
					\includegraphics[width=0.49\textwidth]{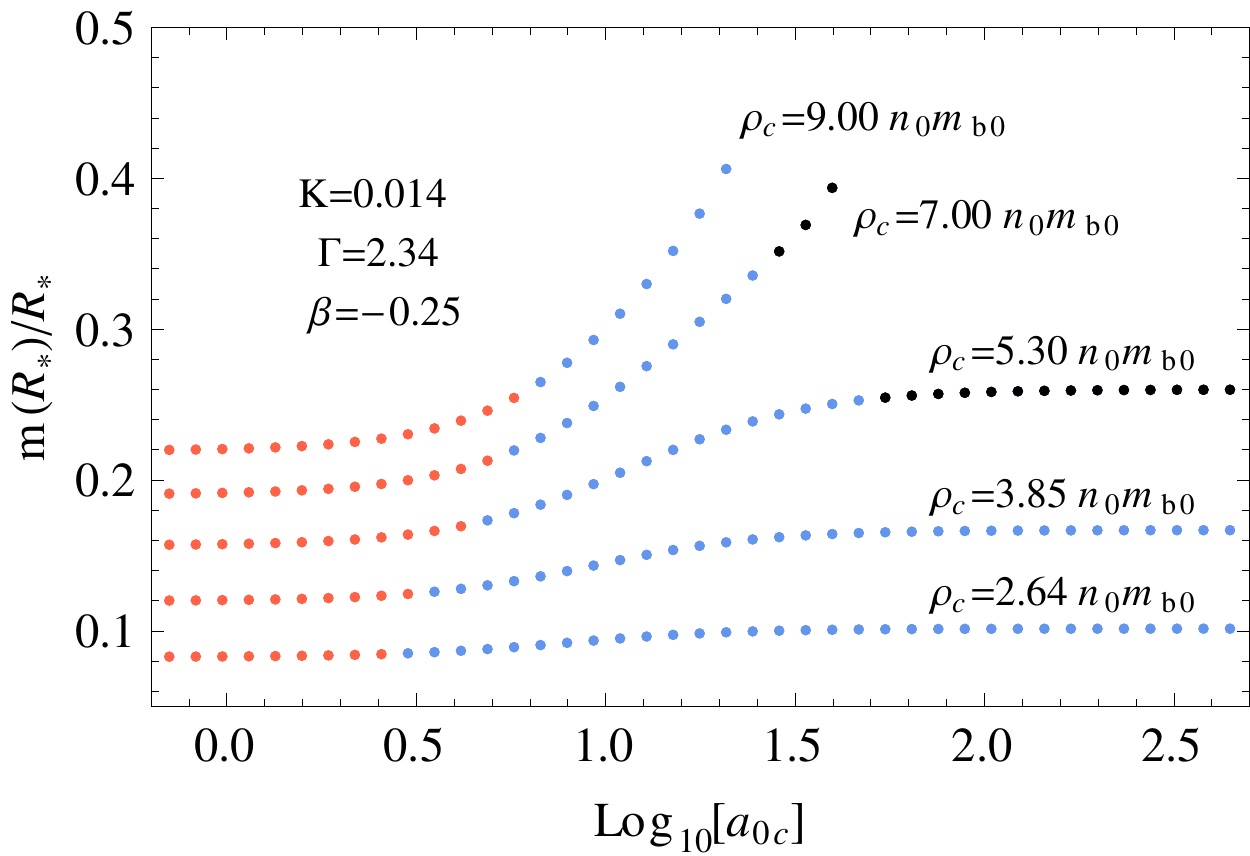}    \ 	\includegraphics[width=0.48\textwidth]{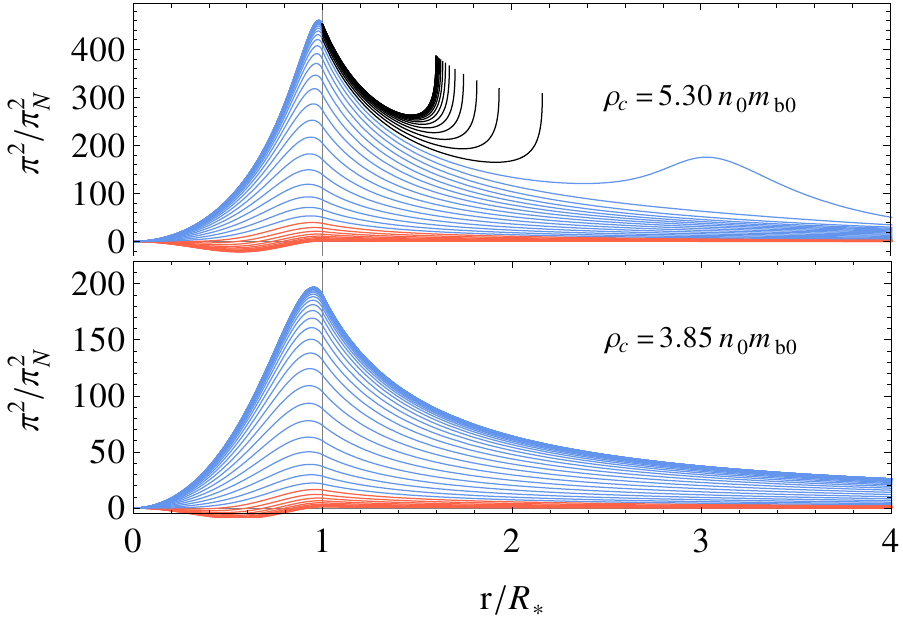}    
					\caption{Range of interest for $a_{0c}$. Each point in the left panel corresponds
                                          to a static configuration  computed with the central density
                                          shown in the plot.  At high densities there is a maximum $a_0$ beyond which star configurations do not exist. Red points represent solutions
                                          for which $\pi$ is complex, blue points regular solutions, and black points solutions that cannot be matched
                                          to a regular exterior solution. The right panel shows the interior and exterior scalar field profiles for configurations
                                          with the same central density, each profile is associated to a point in the left panel.
                                           For
                                          $a_{0c}\lesssim 1$, $m(R_*)$ and $R_*$ remain constant. For $a_{0c}\gtrsim 10^{2}$ and low densities, $m_*$ and
                                          $R_*$ increase with $a_{0c}$, but the ratio
                                          $m_*/R_*$ remains constant.  }
					\label{fig:lowdensityscaling}  
				\end{figure}

			At first sight,  Fig.~\ref{ns-vgbetaminus1}   would seem to suggest    that NS solutions exist for arbitrarily large values of $a_{0c}$. However, this is not 						always the case. In Fig.~\ref{fig:lowdensityscaling}  we explore a larger range of values of $a_{0c}$ for fixed central NS densities
			approximately up to the critical density associated with  maximal mass in the $M$-$R$ curves. We find that the properties of 
			solutions with high $a_{0c}$ depend on $\rho_c$, the value of the density at the center of the star. For low densities, there are equilibrium configurations in the complete range
			of $a_{0c}$ that we explored. Instead,  for relatively high densities we find  an upper bound on $a_{0c}$ beyond which the radius
			of the star cannot be defined in the standard way, since we do not find  any radial position $R_\star$ where the pressure of matter vanishes, hence the star has no natural  boundary.
			
		There exist additional constraints on the allowed range for 
			 $a_{0c}$, which arise when studying in detail  the external field configurations.
			 When a  low central density $\rho_c$ is chosen 
			all the solutions can be matched to healthy exterior solutions. But when  $\rho_c$  is increased, this is not the case. 
			The black points 
			in the left panel of Fig.~\ref{fig:lowdensityscaling} correspond to  interior solutions that can  not be matched to  an asymptotically flat exterior 					spacetime,  the black lines in the right panel show that  the scalar field  profile   $\pi$ diverges for these solutions.
			
			
			
\begin{figure}
                          \includegraphics[width=0.45\textwidth]{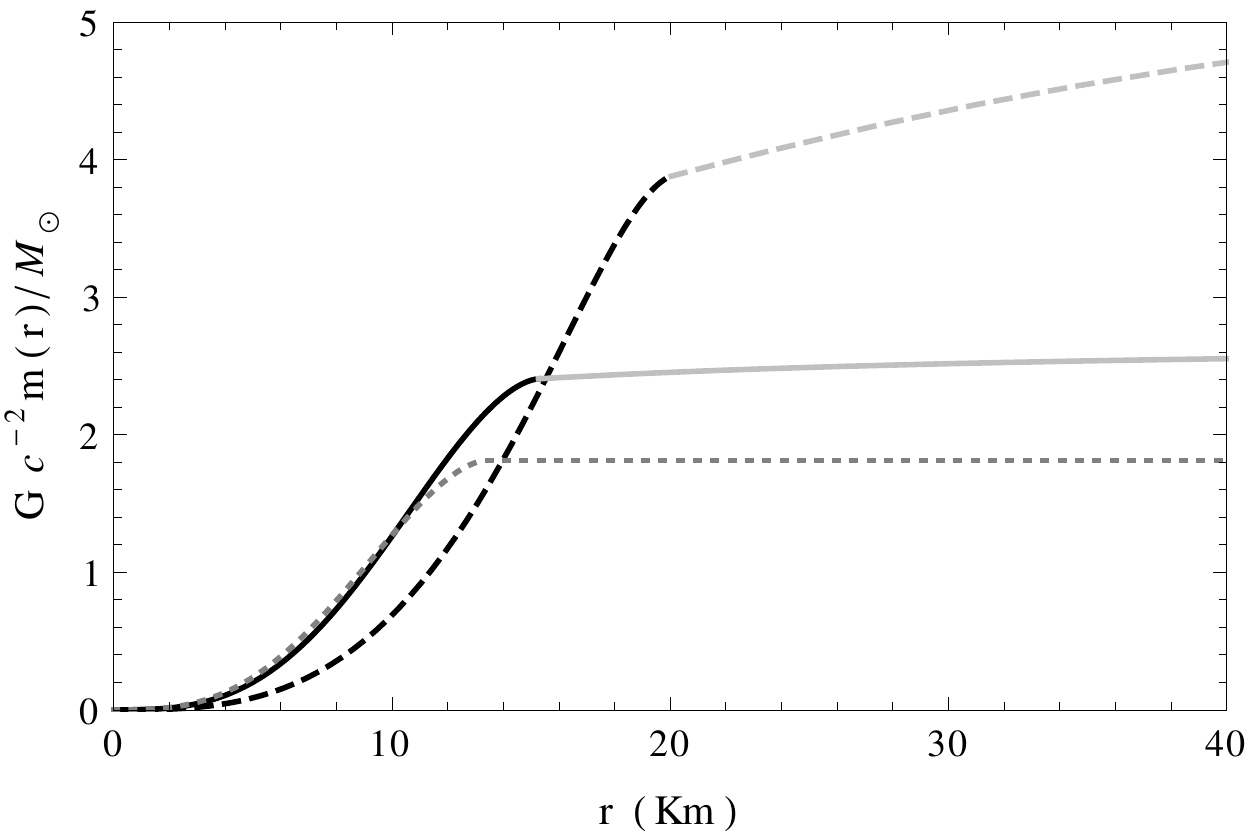}
                          \ \includegraphics[width=0.45\textwidth]{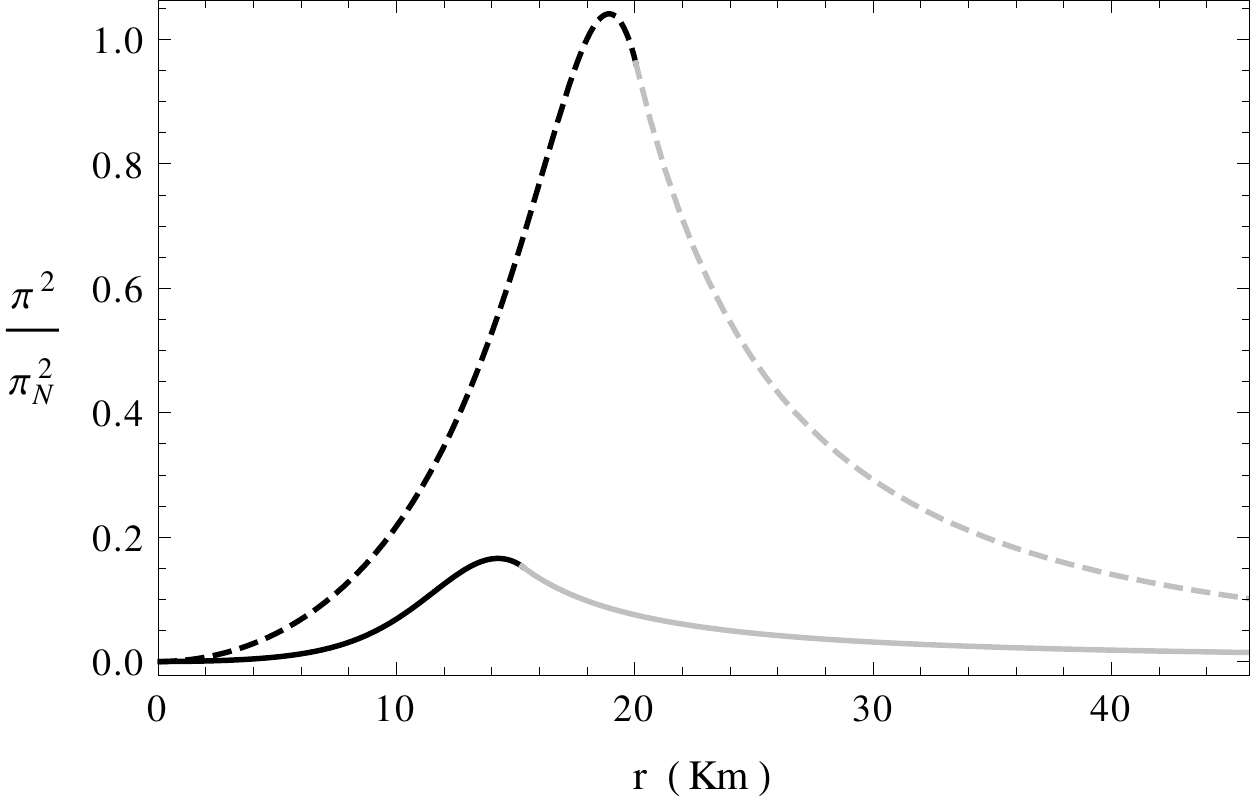}
                          \caption{Interior (black line) and exterior solutions (grey
                            line) for $\beta=-1/4$. Solid lines are for $a_0(0) = 6.0$ and
                            dashed lines for $a_0(0)= 12.0$.  The central density is
                            $\rho_c = 5.3 n_0 m_{b0}$. The left panel shows $m(r)$ in
                            units of solar masses, both lines reach asymptotic constant
                            values. The dotted line is the GR solution. 
                             }
				\label{fig:matchbnl0} 
			\end{figure}

Interestingly, 
			by calculating the NS mass using asymptotic properties of the exterior solutions, we  always
                                find  a NS mass larger than in GR. We   exemplify this fact  in Fig. \ref{fig:matchbnl0}.


\subsection{Neutron star compactness}
As well as in the black hole case, the compactness of neutron stars
is an important
  property  that can be used for observationally characterizing these objects.
  %
  %
  %
In this subsection we show that, despite the constraints one has to satisfy, the
exterior solutions of NS lie in a region of the parameter space where the NS
compactness can be increased with respect to GR.

In GR the exterior solution of a NS is exactly Schwarzshild. The compactness is then
determined by dividing the Schwarzschild mass (in units of distance) over the radius of the
star, and since $R$ is always larger than the Schwarzschild radius the
compactness is always less than that of a Schwarzschild black hole, i.e., $1/2 = M/r_s > M/R$. 

In vector Galileons, as shown in the previous sections, the asymptotic mass gets
contributions from the vector field outside the star which can assume a non-trivial profile. Therefore, in order to obtain the
compactness of a NS we first  extract $M$ from the exterior solution evaluated at a large
$r$ -- where the asymptotic Schwarzschild solution holds -- and then divide it by the radius of
the star. The result is always smaller than the compactness of a vector Galileon black
hole. This is expected,  but interestingly,  the compactness of such black holes can be
larger than that of a Schwarzschild BH, opening the possibility for NS in vector Galileon
theories that are {\it 
 more compact } than GR black holes.

\begin{figure}[h] 
  \includegraphics[width=0.85\textwidth]{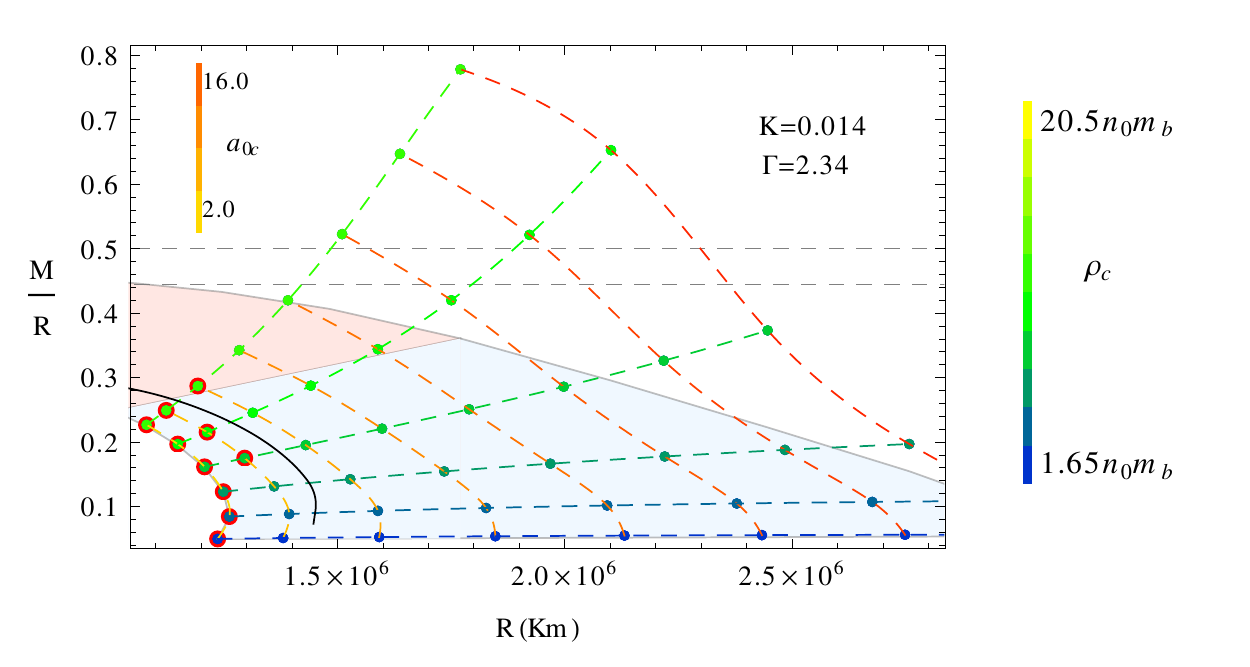}\caption{Compactness of  NS. This quantity is computed with
    the Komar mass of the exterior solutions that match each NS configuration with the
    central value of $\rho$ and $A_0$ indicated by the color bars in the plot. Only stable
    configurations are shown. The black solid line shows the compactness of the same
    configurations in GR for polytropic stars modelled so that the M-R curve gives a
    maximum mass around $2M_\odot$. For the vector Galileons we chose $K=0.014$ and $\Gamma =
    2.34$. The dashed horizontal lines show GR limits for the compactness of spherically
    symmetric and static solutions: $M/R = 0.5$ is the compactness of Schwarzschild
    BH's, and $M/R = 4/9$ is the compactness of incompressible stars.  Notice that, thanks to the properties of our  exterior solution, in some cases NS can be more compact than Schwarzschild black holes. For reference, the blue shaded
    region outlines instead  the compactness of stable NS configurations
     if 
     the exterior solution
    is set to be Schwarzschild geometry; the red shaded part extends this  region to higher densities,
    where the solutions are unstable. These shaded regions show that NS with exterior configurations corresponding to Schwarzschild can not
    be more compact than black holes. }
  \label{bhcvsqfixp}
\end{figure}

\bigskip

In Fig.~\ref{bhcvsqfixp}  we represent the results of a numerical analysis aimed to exemplify the discussion above.
  Each point in the plot corresponds to a
global solution (i.e., a NS plus the exterior solution) with initial conditions indicated by the
colour bars in the plot: the initial condition $\rho_c$ is given by the point colour and
the bar to the right of the plot, while $A_{0c}$ is given by the dashed line on which the point
resides and the colour bar inside the plot. The remaining initial conditions are
$g_{tt}(0)=1$, $g'_{tt}(0) = 0$, and $A_0'(0)  = 0$. The interior of the NS is modelled using a
polytropic EoS with $K = 0.014$ and $\Gamma = 2.34$. This choice leads  for $M$-$R$ curves with
$M_{max}\geq 2M_\odot$ and real $\pi$.  For the solutions circled in red, $\pi$ becomes complex
inside the star. From each interior solution we obtain the radius $R$ of the respective
star, which is given in the $x$-axis of the plot, and a set of initial conditions for the
exterior solution. From the exterior solution we determine $M$. Knowing
$R$ and $M$ we calculate the compactness of each star, and the results are shown in the $y$-axis
of the plot.

Besides the compactness of NS in vector Galileons, Fig.~(\ref{bhcvsqfixp}) also displays
the following reference lines: the dashed horizontal lines show GR limits for the
compactness of spherically symmetric and static solutions: $M/R = 0.5$ is the compactness
of Schwarzschild BH's, and $M/R = 4/9$ is the compactness of incompressible
stars. EoS-dependent computations put lower limits on the compactness of NS, these are not
shown in the plot.  The black solid line shows the compactness of star configurations in
the same range of densities shown in the plot but computed in GR and with polytropic
parameters $K=0.0225$ and $\Gamma=2.34$, so that the M-R curve gives a maximum mass
around $2M_\odot$. The blue shaded region outlines the compactness of the same neutron
stars solutions when we ignore the contributions of $A_\mu$ outside the star, i.e. we
construct the global solution taking the same set of vector Galileon NS configurations
corresponding to the points in the plot, but using a Schwarzschild configuration for the exterior
metric. The red shading extends this to solutions with higher values of the central
density, these solutions are unstable.  As the plot shows, the solutions in the shaded region can
never be more compact than a Schwarzschild black hole; this  option is possible only for vector Galileons, thanks 
to the interactions between gravity and the vector field in the exterior of the star.


\bigskip

 To conclude our discussion on neutron stars, we comment on possible future developments, in case
 the non-minimal  vector-tensor couplings we consider is used to parameterize  modifications
 of Maxwell electromagnetism in strong gravity regimes. While in  
   this section we considered a situation in which matter in the NS interior does not directly couple to the vector (but only indirectly through
 gravity) we could directly couple $A_\mu$ with internal currents.
 It is known that for  standard electromagnetism such couplings can modify the equilibrium configurations \cite{Bekenstein:1971ej}
since electric currents  modify  the equations of hydrostatic equilibrium. A complete  treatment of this topic for vector Galileons is
 under investigation, and we include preliminary results in Appendix \ref{app:currents} . We find that, in certain cases, the inclusion of an interaction term $A_\mu J^\mu$ does not spoil the existence of neutron stars solutions in this model, and can  improve the properties of the scalar field $\pi$, in such a way that configurations with complex  $\pi$  (that we had to discard
 in our analysis above)  
 can be turned into configurations with real scalar $\pi$. Other possible developments left for the future include the addition of rotation and magnetic fields (with $F_{ij}\,\neq\,0$) to explore possible connections with magnestars. We make preliminary steps to find configurations with magnetic fields for vector Galileons   in Appendix \ref{magneticsolutions}, while rotating exterior configurations (for small values of the rotation parameter) are discussed in \cite{Chagoya:2016aar,Minamitsuji:2016ydr}. In the future, we plan to study whether $ I$-$Love$-$Q$ relations \cite{Yagi:2013bca,Yagi:2013awa}  -- which  relate  the moment of inertia $I$, the
 tidal Love number, and the quadrupole moment $Q$ of the star -- get modified in the context of vector Galileons.

					\section{Discussion}\label{sec:disc}

We studied black hole and neutron star configurations in a vector-tensor theory of gravity, a special case of vector
Galileons, described by a simple, one parameter modification of the Einstein-Maxwell action, corresponding to the following non-minimal vector-tensor coupling
$$
\beta\,\sqrt{-g}\,G_{\mu\nu}\,A^\mu\,A^\nu\,,
$$
with $G_{\mu\nu}$ the gravitational Einstein tensor and $\beta$ a coupling constant.
Such coupling term can be used in the contexts of vector inflation, or a vector governing dark energy  or  dark matter. Alternatively, one may think of this vector as a way to parameterise deviations from the standard Maxwell's electromagnetism in regimes of strong gravity.   The physics
of black holes reveal surprisingly rich properties, that generalize  the standard Reissner-Nordstr\"om solution
of Einstein-Maxwell theory of gravity, but also black hole configurations of related scalar-tensor theories. There exist
two disconnected branches of static, spherically symmetric, asymptotically flat solutions, whose features we studied in detail. Solutions have a mass $M$ and a vector charge $Q$, and are also characterized by an integration constant $P$ mainly controlling the profile of the  longitudinal scalar polarisation of the vector ($P$ is not however associated to a conserved asymptotic charge). The existence and position of horizons depend in a non-trivial way by the parameters involved. In some cases analytic configurations are available, while for most values of the parameters involved we extract the behaviour of solutions through a numerical analysis. We find that in certain regions of parameter space black holes can be  more compact than in General Relativity, providing a  distinctive feature of black holes in this vector-tensor set-up. In an Appendix, moreover, we study in detail differences and similarities of our black hole configurations with solutions of  scalar-tensor theories.


With respect to the neutron stars, our study shows that the vector profile plays an important role in determining the star configuration, both for controlling its internal hydrostatic equilibrium configuration, and for determining the external gravitational solution which is generally not described by a pure Schwarzschild geometry. The properties of neutron stars in this vector-tensor theory  
are quite rich; they can be larger and more massive  than their GR counterparts 
and, for certain parameter choices, more compact. In some cases, they might be even more compact than Schwarzschild black holes, making this objects observationally interesting for the prospect of gravity wave detection. We also comment on possible generalizations to magnetically charged or rotating configurations, presenting in appendixes some preliminary calculations on this respect. This can be of interest  especially when this vector-tensor non-minimal coupling is used to describe deviations from Maxwell theory, for describing exotic compact objects as magnetars. 

 Future interesting developments will include a study of rotating exterior solutions for neutron stars and black holes -- with arbitrarily large rotation parameter -- and an analysis of possible  generalizations of I-Love-Q relations for neutron stars in vector Galileon theories. We hope to report soon on these topics.

\section*{Acknowledgments}
 JC is supported by STFC and CONACyt grants 263819 and 179208.
GN is supported by CONACyT grants: 179208, 269652, Fronteras de la Ciencia 281, and by DAIP-Universidad de Guanajuato grant 1,046/2016.

\bigskip
\section*{Appendixes}
\begin{appendix}

\section{Black hole configurations for the  branch  $\pi=0$}\label{app:pi0}

In this Appendix we discuss black hole configurations in the second branch of solutions satisfying the constraint \eqref{constraint} with $\pi=0$.
This branch contains the Reissner-N\"ordstrom configuration in the limit $\beta=0$. However, for non-vanishing $\beta$, the profile for $A_0$  changes
the geometry more drastically than in the other branch  $\pi\neq0$, leading to modifications of the GR geometry at large distances.


The algebraic constraint \eref{constraint} vanishes identically for this branch, and the equations of motion \eref{eeh}-\eref{veh} give two independent differential equations for $A_0$ and one of the metric functions $n$ and $m$. It is, however, more convenient to present the system as a set of three differential equations that
are first order in derivatives of metric, namely
   \begin{subequations}
   	\begin{align}
   	0 = & r^2 \kappa  (r-2 m{}) {A_0}'{}^2-4 \left(r-r \beta  \kappa  {A_0}{}^2-2 n{}\right) m'{} , \\
   	0 = & {}{A_0}'{} \left[r \left(n{} \left(-5+2 m'{}\right)+r \left(2-m'{}+n'{}\right)\right)+m{} \left(8 n{}-r \left(3+2 n'{}\right)\right)\right] \nonumber \\
   	& +(r-2 n{}) \left(-4 \beta  {}{A_0}{} m'{}+r (r-2 m{}) {}{A_0}''{}\right), \\
   	0 = & 8 n{}^2+2 r n{} \left(-2+2 \beta  \kappa  {}{A_0}{}^2+8 r \beta  \kappa  {}{A_0}{} {}{A_0}'{}+r^2 \kappa  {}{A_0}'{}^2-4 n'{}\right) \nonumber \\ 
   	& -r^2 \left(8 r \beta  \kappa  {}{A_0}{} {}{A_0}'{}+r^2 \kappa  {}{A_0}'{}^2-4 n'{}+4 \beta  \kappa  {}{A_0}{}^2 n'{}\right)+2 m{} \left[8 r \beta  \kappa  {}{A_0}{} (r-2 n{}) {}{A_0}'{} \right. \nonumber \\ 
   	& \left.+(r-2 n{}) \left(2+r^2 \kappa  {}{A_0}'{}^2-4 n'{}\right)+2 \beta  \kappa  {}{A_0}{}^2 \left(r-4 n{}+2 r n'{}\right)\right].
   	\end{align}
   \end{subequations}
  The first (third) equation is algebraic in $n$ ($m$), and the result can be plugged back in the remaining 
  two equations to find the two independent, but not first order in derivatives of the metric functions, equations
  for $m$ ($n$) and $A_0$.
  




We can not determine
an analytical solution for this branch. We can proceed with studying  solutions using  perturbative expansions, as
done for the first branch $\pi\neq0$ in the main text. 


\subsubsection*{Asymptotic expansion for large values of the radial coordinate }
For $r\gg 1$ we 
apply   the same perturbative approximations introduced in  \eref{g1fep}-\eref{g1a0ep}, and we make an expansion in a small $1/r$ regime for $n(r)$, 
 $A_0(r)$, and $m(r)$.
 We find
\begin{subequations}
	\begin{align}
	{} 1-\frac{2\, n(r)}{r} = & 1-\epsilon \frac{2 [M+\kappa\beta  P (M P+2 Q)  ]}{r \left(1-P^2 \beta  \kappa \right)} +\frac{\epsilon^2 \kappa}{2r^2\left(1-P^2 \beta \kappa \right)^3 } \Big\{ Q^2 (1-4 \beta )  \label{npi0asymp}\nonumber \\ 
	{} &  - \kappa P^2 \beta  \left[Q^2+4\left(2 M^2 P^2+2 M P Q- Q^2\right) \beta\right]  \nonumber \\ 
	{} &  -8 \kappa^2 P^4 \beta^3(M P+Q) (M P+2 Q) \Big\}  + \mathcal O(\epsilon^3)\, , \\
	{} 1-\frac{2\, m(r)}{r} = & 1-\frac{2 M \epsilon }{r}+\frac{\epsilon^2 Q^2 \kappa }{2 r^2(1 - \kappa \beta P^2) } + \mathcal O(\epsilon^3) \, , \\
	{} \hskip 2.2em A_0(r) = & P+\frac{Q \epsilon }{r}-\frac{P Q (2 M P+Q) \beta  \epsilon ^2 \kappa }{2 r^2 \left(1-\kappa \beta  P^2 \right)} + \mathcal O(\epsilon^3)\, . \label{api0asymp} 
	\end{align}
	\end{subequations}
In contrast to the case $\pi\neq 0$,  the asymptotic geometry in this branch is more sensitive  to the vector 
charge, which affects the geometry contributing   already at first order in  the expansion parameter $\epsilon$.
It would be interesting to study in  detail whether asymptotic scalar  
  charges can characterise  black hole configurations in this branch. Notice however that we could define  a {\em stronger}
weak-field limit, imposing that the parameter  $P$, which enters at $r^0$ order in the expansion for $A_0$, is small:
 $P\to \epsilon\, P$.
  In this case,  the  corrections to the geometry associated with the vector contributions (parameters $P$ and $Q$) are  pushed to second order in an $\epsilon$ expansion, and the geometry would be characterized by a Komar mass controlled 
  by the parameter $M$. 
  

 
 
\subsubsection*{Small $\beta$ expansion }

This branch is smoothly connected to the RN solution in the limit $\beta\to 0$. The simplest way to see this is to consider the following
ansatz:
\begin{align}
1-\frac{2 n(r)}{r} & = 1- \frac{2 M}{r} + \beta f_1(r) + \beta^2 f_2(r) + \dots, \nonumber \\
1- \frac{2m(r)}{r} & = 1 - \frac{2 M}{r} + \beta g_1(r) + \beta^2 g_2(r) + \dots, \nonumber \\
A_0(r) & = \sqrt{\beta} a_1(r) + \beta^{3/2} a_2(r) + \dots ,\\
 \end{align}
and solve for the functions $f_1, f_2, g_1, g_2, a_1, a_2, \dots$. At leading order in $\beta$ the RN solution is generated:
\begin{align}
g_1 = f_1 & = \frac{Q^2}{4 M_p^2 r^2}, \\
a_1 & = P - \frac{Q}{r}.
\end{align}
This leading order solution perturbatively reconstruct the first term in the RN geometry. 
At next-to-leading order we find corrections  only to the time component of the metric:
\begin{align}
g_2 & = a_2  = 0, \\
f_2 & = -\frac{2 M P^2}{M_p^2 r}+\frac{2 P Q}{{M_p}^2 r}-\frac{Q^2}{{M_p}^2 r^2}.
\end{align} 
It is possible to solve to higher orders in $\beta$, where corrections to $A_0$ and $g_{rr}$ are present too, however the analytic 
expressions are not particularly interesting. We verified that solving to the next orders in $\beta$ does not lead to new integration
constants: any integration constants that arise at each higher order, can be reabsorbed in the definitions of $M$, $P$ and $Q$.


\section{Exterior solutions in the branch $\pi\neq0$} \label{sec:numext}
			In this appendix we numerically
			study  exterior solutions for neutron star configurations, for different values of the parameter $\beta$ in the branch $\pi \neq 0$.
		Although we have in mind neutron stars, the same analysis remains valid outside the horizon of black hole configurations.   
Hence			
			we study  numerical solutions
			to
			eqs. (\ref{eq:00met},\ref{eq:0vec}) for $A_0(r)$ and $n(r)$
			for representative values of  $\beta$, and use
			eqs. (\ref{eq:mofn},\ref{eq:pisqr1}) to determine $\pi(r)$ and $m(r)$. We use c.g.s. units to ease the comparison to the literature on NS. To find numerical solutions, 
			we set initial conditions at a typical radius of neutron stars, $R_*=
			12$ Km, and take values for $n(R_*)$, $n'(R_*)$, $A_0(R_*)$ and $A_0'(R_*)$ within a range
			motivated 
			by
			the interior numerical solutions that we  investigate in the main text.
			From now on, all quantities with a subscript `$*$' denote quantities
			evaluated at $R_*$, for example $A_{0*}=A_0(R_*)$.
			In order to work with  dimensionless 
			quantities, we introduce a normalisation  for $A_0$:   $a_0 = A_0/
			(10^{23} {\rm cm}^{1/2}{\rm gr}^{1/2}/{\rm s})$, with $a_0$ dimensionless.  This 
			choice is made to match the 
			order of magnitude of the quantity
			$1/\sqrt{\kappa} = 6.94\times 10^{23} {\rm cm}^{1/2}{\rm gr}^{1/2}/{\rm s}$ which turns out to have the same dimensions. 
			Analogously, we also include a normalization for the scalar, $\pi_N =10^{23} {\rm cm}^{1/2}{\rm gr}^{1/2}/{\rm s}$, and
			plot the dimensionless  quantity $\pi/\pi_N$.
			\begin{figure}[!h!]
				\hspace{-1em} \includegraphics[width=0.46\textwidth]{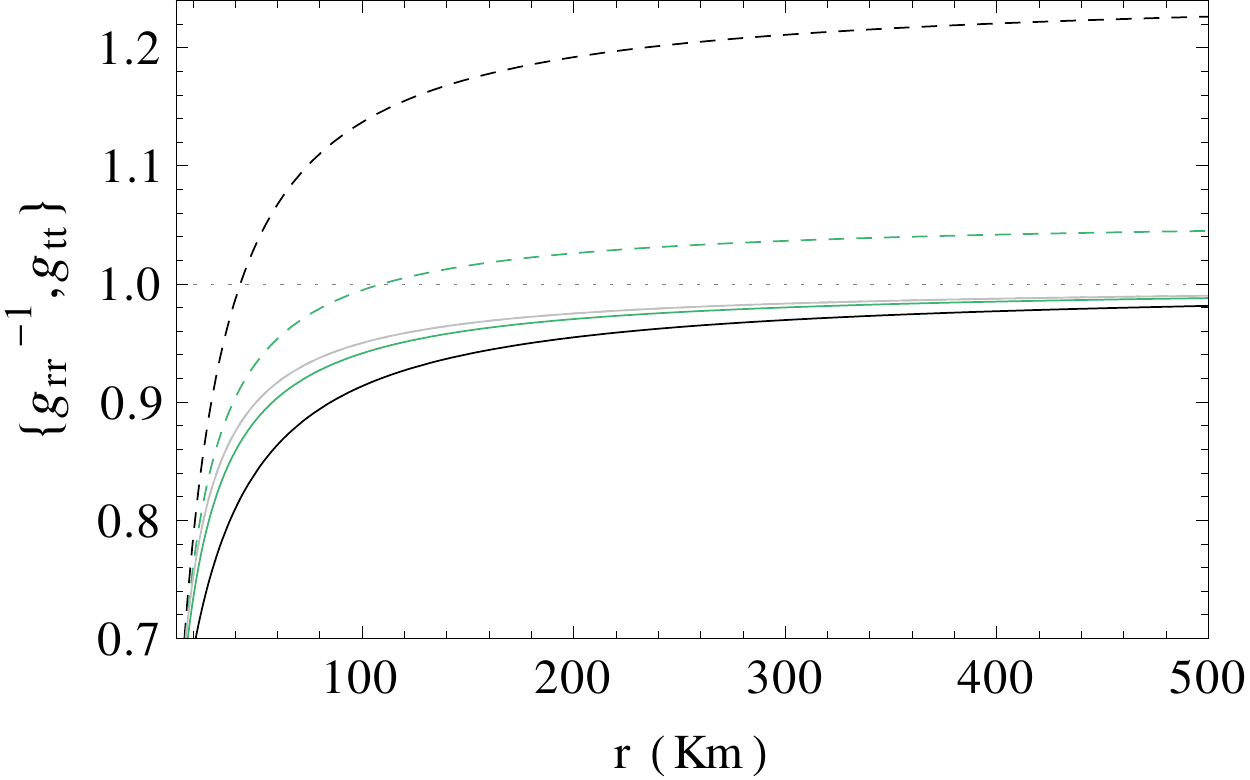} \ \includegraphics[width=0.45\textwidth]{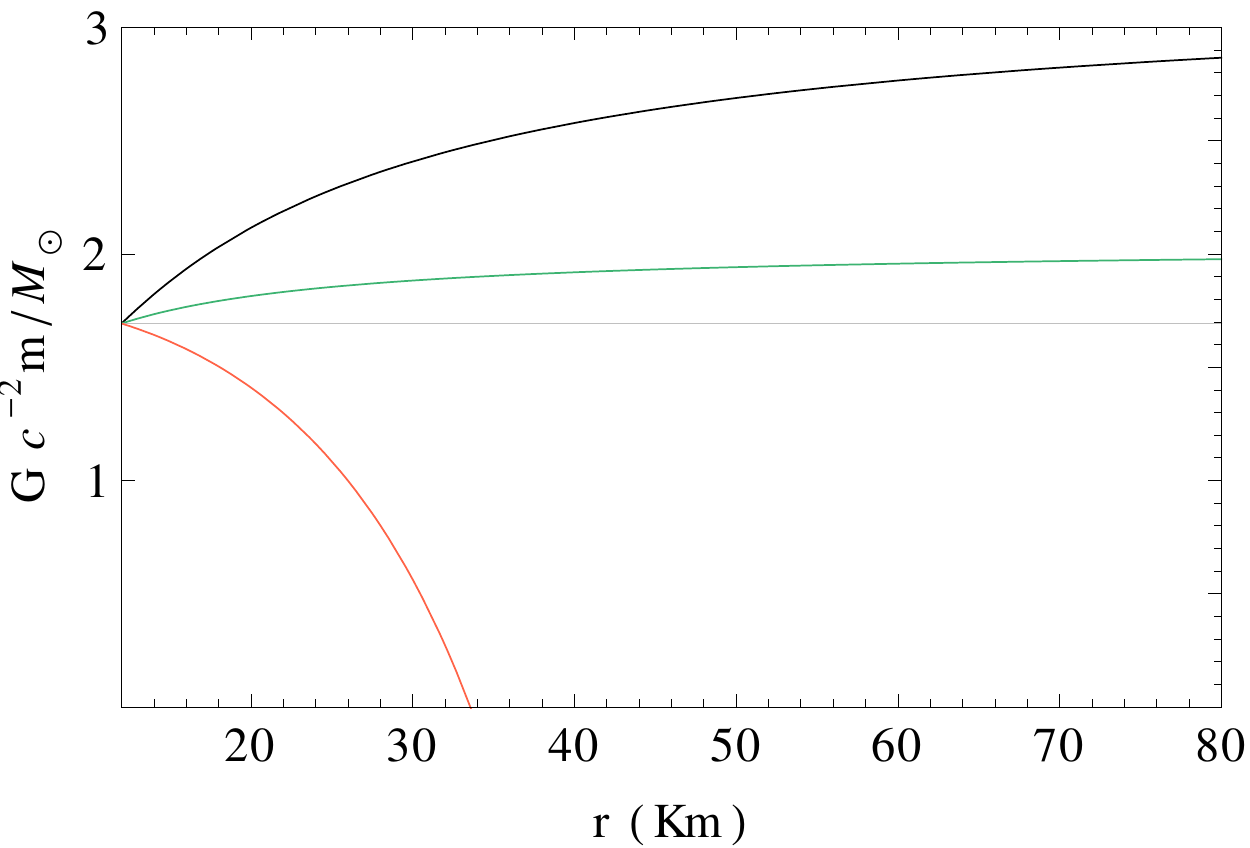} \\ 
				\includegraphics[width=0.45\textwidth]{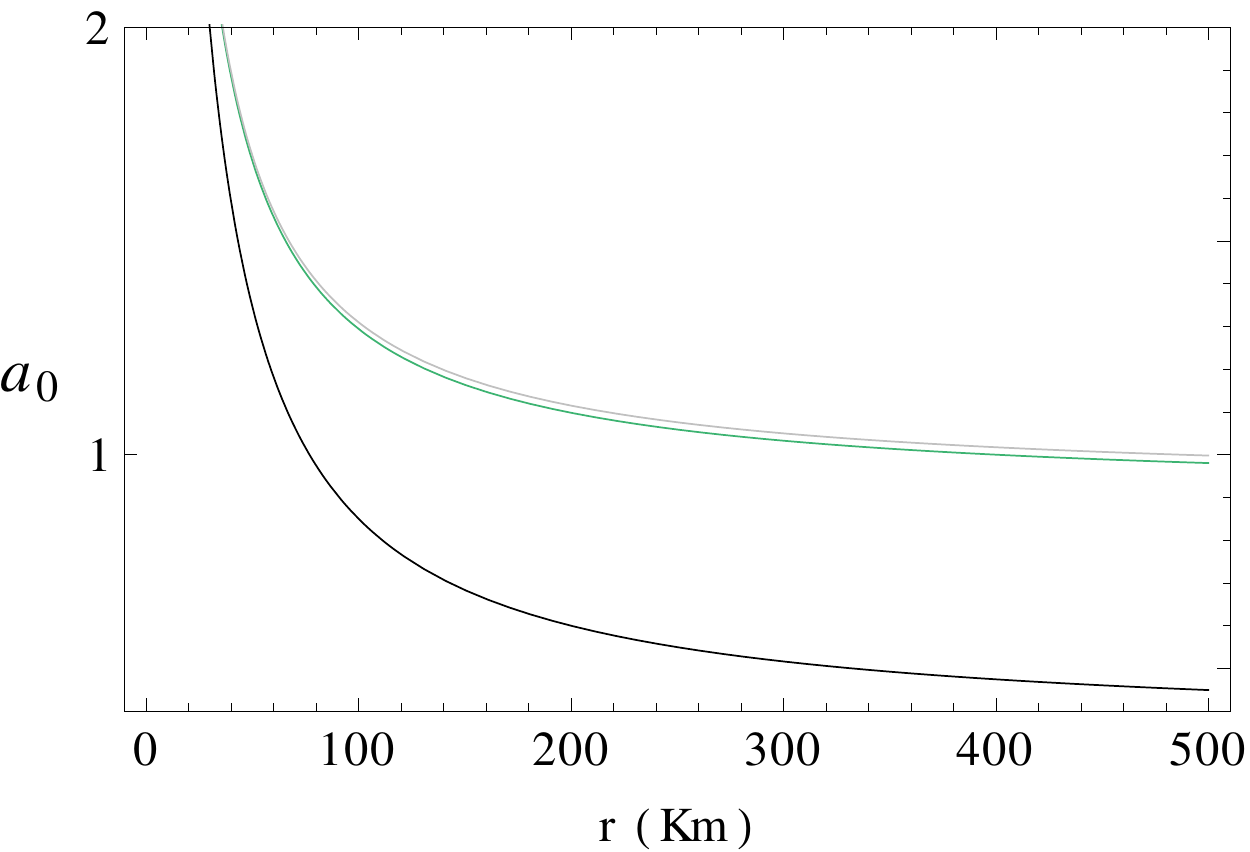}
				\ \includegraphics[width=0.475\textwidth]{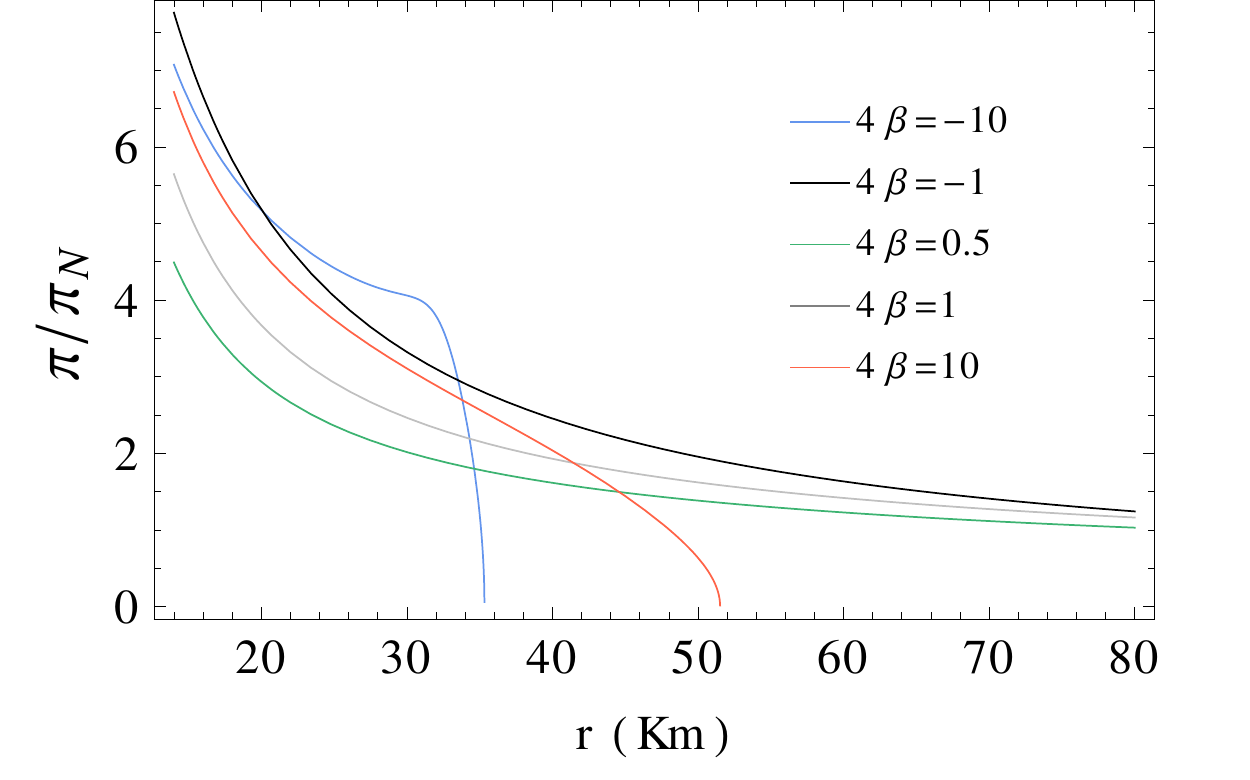}\caption{ Clockwise from
					top-left, we represent various quantities as functions of radius $r$: a) $g_{tt}$(dashed lines) and $g_{rr}^{-1}$ (continuus line) components of the exterior metric; b) 
					$m(r)$ measured in solar masses; c)  $a_0(r)\,=\,A_0(r)/  (10^{23} {\rm cm}^{1/2}{\rm gr}^{1/2}/{\rm s})$; d)  the dimensionless quantity $|\pi(r)/\pi_N|$. The colour code in
					each plot is  indicated in the bottom right panel. 
					The horizontal line in the top left panel a) is a reference to make it easier to verify visually that $g^{rr}\to 1$ asymptotically. The
					initial conditions are $n_* = 2.5{\rm\ Km}, n'_*=0, a_{0*} = 4, a'_{0*} =-0.27{\rm\
						Km}^{-1}$. See the main text for an analysis of the plots. 
				}
				\label{grmext}
			\end{figure}
			
			We plot our results for a representative choice of parameters in the Figure \ref{grmext}, and comment on their physical consequences below.
			\begin{itemize}
				
				\item Fig.~\ref{grmext}a represents the metric components $g_{rr}^{-1}$ and $g_{tt}$ as function of the radial coordinates,
				for different choices of $\beta$ (the colour code for each line is explained at the bottom right panel of Fig.~\ref{grmext}). When $\beta\neq1/4$, these quantities are not equal. Asympototically, $g_{rr} \to 1$ in all cases, while in general $g_{tt}$ does not tend to unity
				at large $r$.
				Nevertheless, we can always redefine the time coordinate
				in such a way that  $g_{tt} = 1$ for large $r$, so to have an asymptotic configuration corresponding to Minkowski space written in spherical coordinates.
				
				\item Fig.~\ref{grmext}b  represents the radial evolution of   mass of the object in units of Solar masses, for different choices of $\beta$. The asymptotic values of these quantity correspond  to the Komar mass. While for $\beta=1/4$ the  function $m(r)$ stops evolving
				outside the star, for different $\beta$'s the vector field continues to source the Einstein tensor and consequently the
				gravitational mass is not fixed at $R_*$. As a result, the  mass observed
				asymptotically in the models with $\beta<1/4$ is always larger than in GR. The models with
				large positive $\beta$ have to be discarded since $m$ takes negative asymptotic values. Indeed, we have checked that for large values of $\beta$
				the solutions develop essential singularities at large values of $r$ outside the star surface, hence they are physically not interesting. 
				
				\item Fig.~\ref{grmext}c  represents the radial profile of the time-like component of the vector field, $a_0=A_0/(10^{23} {\rm cm}^{1/2}{\rm gr}^{1/2}/{\rm s})$.  
				For any $\beta\neq1/4$, $A_0$ decays faster
				than for $\beta=1/4$ at large $r$; that is, it decays faster than $1/r$. 
				  We do not find any hint of singular behavior in the numerical solutions for $A_0$.  
				
				\item Fig.~\ref{grmext}d represents the radial profile for the absolute value of the scalar field $|\pi|$.  While for small values of $\beta$ such radial profile is smooth, it develops 
				singularities when $\beta$ is sufficiently large (of order 10 in absolute value), since such function becomes negative at finite values of $r$.  This should be connected with the singularities in the geometry  that we find for large values of $\beta$. 
				
			\end{itemize}

			The results obtained indicate that $|\beta|$ cannot be arbitrarily large, otherwise the scalar field becomes complex.			

 \section{Difference between vector  Galileon and scalar Horndenski theories}\label{sec:diff}
 
 In this appendix
we   discuss how black hole configurations   in 
   the vector-tensor (VT) system \eref{eq:action} we are considering differ from  scalar-tensor (ST) theories as Horndeski. 
   A comparison
between these theories is not straightforward given the distinct nature of the fields involved. For example, in general
 $\nabla_\mu A_\nu\neq \nabla_\nu A_\mu$ unless $A_\mu = \partial_\mu \phi$ for some
scalar field $\phi$.
 Furthermore, 
the number of equations of motion is in general different in the vector and scalar cases. The main message of this
appendix
    will be  that, when a comparison is possible, the vector Galileon model allows for physically interesting generalisations of the
black hole solutions found in the Horndenski scalar case, also clarifying some of the properties of the latter. 

\subsubsection*{First difference: the set of equations}

Black hole configurations have been extensively studied in scalar-tensor theories. A particularly 
interesting subclass of scalar-tensor theories is known as the \emph{Fab Four} \cite{Charmousis:2011bf}, and has the property that it is the only subset of
Horndeski's gravity where  the cosmological constant can be self-tuned, meaning that Minkowski and cosmological solutions are allowed
for any value of the bare cosmological constant. Within the Fab Four, there are minimal subsets that allow for self-tuning solutions
while keeping some simplicity in their analytical treatment. An example is the action
\begin{equation}
S = \int d^4 x\sqrt{-g} \left[ \tilde \kappa (R - 2 \Lambda) - \frac{1}{2} \left(  \tilde \alpha g^{\mu\nu} - \eta G^{\mu\nu}  \right) \nabla_\mu \phi \nabla_\nu \phi   \right] + S_m, \label{john}
\end{equation}
where $\tilde \kappa$, $\tilde\alpha$ and $\eta$ are coupling constants and $\phi$ a real scalar field.
 in the language of Fab Four, this     corresponds to a combination of the \emph{George} and \emph{John} Lagrangians that has been
studied in cosmological and static contexts (see for example \cite{Bruneton:2012zk,Maselli:2016gxk}). When focussing on static spacetimes,  the choice $\tilde \alpha = 0$ and  a scalar field profile with linear time dependence 
\begin{equation}
\phi(t,r) = P\, t + \psi(r) \label{ansatzst}
\end{equation}
 leads to a black hole solution with a 
 Schwarzschild metric.

 Eq. \eref{ansatzst} is our starting point for a concrete comparison to the vector-tensor model: if we write $A_\mu = \partial_\mu \phi$, then $\partial_r\phi(t,r) = \partial_r\psi(r)$ plays the role
 of $\pi(r)$, and $\partial_t \phi(t,r) = P$ plays the role of a constant time like component   $A_0$ of the vector field. 

At first sight,
one might think that under these identifications the vector system \eref{eq:action} is equivalent to \eref{john} with $\tilde \alpha=0$ (remember that $F_{\mu\nu}$ vanishes for a constant $A_0 = P$). However, this is not the case because the vector equation of motion derived from \eref{eq:action} is more restrictive than the scalar equation derived from $\eref{john}$: $\nabla_\mu (G^{\mu\nu}\partial_\nu \phi) = 0$, where we set $\tilde \alpha = 0$. While in the ST case only the vanishing of the radial
component of $G^{\mu\nu}\partial_\nu \phi$ is required (by the $(t,r)$ component of the metric equations), in the VT case
the time component of $G^{\mu\nu}\partial_\nu \phi$ is required to vanish as well, by the vector equations of motion. Under the 
spherically symmetric ansatz \eref{eq:metric}, and with $A_0 = P$, this last condition is  $P\, m'(r) = 0$. Assuming $P\neq 0$,
the only option left is $m(r)=$constant, which leads to the \emph{stealthy} Schwarzschild solution when in vacuum (see
Section \ref{sec:analytic}). 
 Hence in vacuum, and with the simplest, constant  profile for the vector $A_0$ component, ST and VT have the stealth Schwarzschild
 solution in common. The difference among the two systems  arise in presence of matter: while in the ST case solutions with  ansatz
\eref{ansatzst} can be found, in the VT  with $A_0 = P \neq 0$ the constraint $m' = 0$  forbids the existence of regular solutions.

Therefore, in the specific case $A_0 = P$ the vector model is more restricted than the scalar model. On the other hand, 
the advantage of the VT model is that, as shown in the previous sections, we can turn on a non-trivial, radial dependent  profile for $A_0$ and still find 
physically interesting vacuum solutions, including Schwarzschild, as well as physical solutions in presence of matter
(as discussed in our analysis of neutron star configurations). These solutions
do not have an equivalent in the scalar case:
a static vector field with non-trivial profiles for the time and radial components is not obtainable in a ST theory (see also \cite{Minamitsuji:2016ydr} for a related discussion). 

\subsubsection*{Second difference: black hole hairs}

Another
 physically relevant difference with respect to scalar-tensor theories lies in the ``hair'' of the spacetime solutions. In ST, there is a no-hair theorem for the shift-symmetric sub-sector of Horndeski gravity \cite{Hui:2012qt}, a sub-sector that includes action \eref{john}. Under the assumptions that the spacetime is static, spherically symmetric and 
asymptotically flat, and the scalar field is a function only of $r$,  regular BH solutions do not support non-trivial
profiles of the scalar field  (see \cite{Maselli:2016gxk,Herdeiro:2015waa} for a concise review of the theorem and
 possible ways to 
circumvent it). Relaxing these assumptions BH's with scalar hair have been found \cite{Babichev:2013cya}, although this hair is dubbed 
``secondary'' since it is not independent of the other charges (mass and electric charge for static BH's). Solutions with primary hair
have been found in bi-scalar extensions of Horndeski gravity \cite{Charmousis:2014zaa}. 

It is worth to
clarify whether these no-hair theorems have any consequence for the vector model studied in this paper.  
When $A_\mu = \partial_\mu\phi$ the answer is yes, and the only BH that we can obtain is Schwarzschild. 
This is consistent with ST theories of the form \eref{john} where, as  discussed earlier,
the only solution for $\tilde \alpha=0$ with a scalar field linear in $t$ is the Schwarzschild metric.
 As soon as we allow for a non-trivial profile of $A_0$, 
we break the relation $A_\mu = \partial_\mu \phi$ and we are able to find asymptotically flat BH solutions with primary hair, without
introducing a time dependence in the vector field.  This corresponds to the vector charge, denoted with $Q$ in the main text, analogously to Reissner-Nordstr\"om black holes.
 Recall that  on the  branch of solutions with a non-trivial profile for a scalar field $\pi \neq 0$, we have  an independent integration 
 constant (denoted with $P$ in the main text) which plays an important role in controlling the scalar profile.  
 However, this integration constant is not associated with any conserved 
charge, and does not obey a Gauss law: hence it can not be considered a scalar hair. 




\subsubsection*{Another class of vector Galileon black hole solutions} \label{app-another}
For the sake of further comparison with the scalar tensor case, we can investigate what happens removing the  term $F_{\mu\nu} F^{\mu\nu}$ in \eref{eq:action}. This corresponds to a system that we do {\it not} consider in the rest of the paper,
 but which has the same Lagrangian structure as \eref{john} with $\tilde \alpha = 0$. 
In vacuum, for $A_0 = P$ we have the stealth Schwarzschild solution. 
  A more general solution with non-trivial $A_0$ exists, 
 given by
\begin{subequations}
\begin{align}
			n(r) &= m(r)  = 2M, \\
			A_0(r)^2 & = P^2+\frac{2 P Q}{r}, \\ 
			\pi(r)^2 & = \frac{2 P (M P + Q) r   }{(r - 2 M)^2}.
\end{align}
\end{subequations}

			Asymptotically, the vector field profiles are
			\begin{subequations}
			\begin{align}
			A_0(r) & = P+\frac{Q}{r}-\frac{Q^2}{2 P r^2} + \mathcal O(r^{-3}), \\ 
			\pi(r) & = \sqrt{2 P (M P + Q)}\left(  \frac{1}{r^{1/2}} + \frac{2 M}{r^{3/2}} + \mathcal O(r^{-5/2}) \right).
			\end{align}
			\end{subequations}
			The origin of this solution is again in the vector equation of motion, which  restricts the geometry to be Schwarzschild; and
			in the absence of the restriction  $A_\mu = \partial_\mu \phi$ for some scalar field $\phi$.
			

\section{Charged black hole solutions for $\beta \approx 1/4$}\label{app-ch-sec}

\smallskip
\noindent
In this Appendix we provide more details on the arguments sketched in Section \ref{sec:appro}. 
We examine a
  region of interest in the available parameter space  nearby the value  $\beta = 1/4$ in the branch $\pi\neq 0$. 
To search for solutions in this region,  we express our fields as $n(r) = n_{0} + (\beta - 1/4) n_{1}(r) + \dots$, and
similarly for $A_0$, $m$; then we solve the equations \eref{eq:00met}-\eref{eq:0vec} order by order in $\beta-1/4$.  We find
\begin{subequations}
\begin{align}
{} 1-\frac{2\,n(r)}{r} &  {} \hspace{0em} = 1-\frac{2 M}{r} + (\beta-1/4)\frac{8 Q^2 \kappa}{r^2 (\kappa P^2  - 4)} + \mathcal O[(\beta-1/4)^2]\, , \label{eq:gttnearquarter}\\
{}  1-\frac{2\,m(r)}{r} &    {} \hspace{0em}   = 1-\frac{2 M}{r} + (\beta-1/4)\left( 1-\frac{M}{r}   \right)\frac{16 Q^2\kappa}{r^2 (\kappa P^2  - 4)} + \mathcal O[(\beta-1/4)^2]\, , \label{eq:grrnearquarter}\\
{}  \hspace{2.em}A_0(r) &   {} \hspace{0em}    = P + \frac{Q}{r} + (\beta-1/4) \frac{4 P Q^2 \kappa}{r^2 (\kappa P^2 - 4)} + \mathcal O[(\beta-1/4)^2]\, ,  \label{eq:a0nearquarter}
\end{align}
\end{subequations}
$\pi$ can be determined from \eref{eq:pisqr1}, resulting in the same expression as \eref{g1pi}, plus $\mathcal O(\beta-1/4)$-corrections. 

In order to calculate reliably the position of the horizons $r_h$, located at the points where $g^{rr} = 0$, 
we need to investigate whether these configurations can be trusted along the entire range of radial direction. To do so,
we can estimate the size of higher order corrections $\mathcal O[(\beta-1/4)^n]$, $n\ge 2$, to the previous formulae. 
 The corrections
of order $\mathcal O[(\beta-1/4)^2]$ are not difficult to calculate, and either by numerical analysis, or analytically but using some approximations ($P$ and $M$ small). We find 
  that for some values of the parameters $M, P$ and $Q$ the terms quadratic in
$(\beta - 1/4)$ become
larger than  terms linear on this quantity: there exists a critical value $r_c$ of the radial coordinate below which our expansion
 becomes unreliable. If $r_h\le r_c$, our estimate of the position of horizons is unreliable. 
 A full
expression for $r_c$ in \eref{eq:grrnearquarter} is complicated and not  particularly illuminating.  On the other hand,
 we can get  intuition of its structure by considering  the  simplified case $P=0$ and $M/Q << 1$. Then  we find that our
  series expansion is well defined up to second order in $(\beta-1/4)$,  
  if
      \begin{equation}
   \label{ineco} 3 |\beta-1/4|A_0(r)^2 \kappa \ll 1\,.
   \end{equation}
     We then find an event horizon at the position
  \begin{equation} \label{posh1app}
r_{h} \approx 2 M + (\beta - 1/4)\frac{ 4 \kappa Q^2}{4 M - \kappa M P^2  } + \mathcal O[(\beta - 1/4)^2]\, . 
\end{equation}
as long as our  choice of charges and parameters involved ensures that inequality (\ref{ineco})  
  is satisfied at $r=r_h$. Expression (\ref{posh1app}) shows that the position of the horizons get indeed shifted when $\beta \neq 1/4$, by an amount depending also on the charges involved. 
{Although there is no guarantee that the third and higher order terms in the series expansion \eref{eq:gttnearquarter}-\eref{eq:a0nearquarter} are also consistent under the conditions discussed above, we numerically checked
   that these findings are correct.}

\section{Black hole numerical solutions: the behavior  of the  charges  }\label{app-difpan}

\smallskip
\noindent

\begin{figure}[ht!]  
  \includegraphics[width=0.45\textwidth]{figs/inics.pdf} \hspace{1em}
  \includegraphics[width=0.45\textwidth]{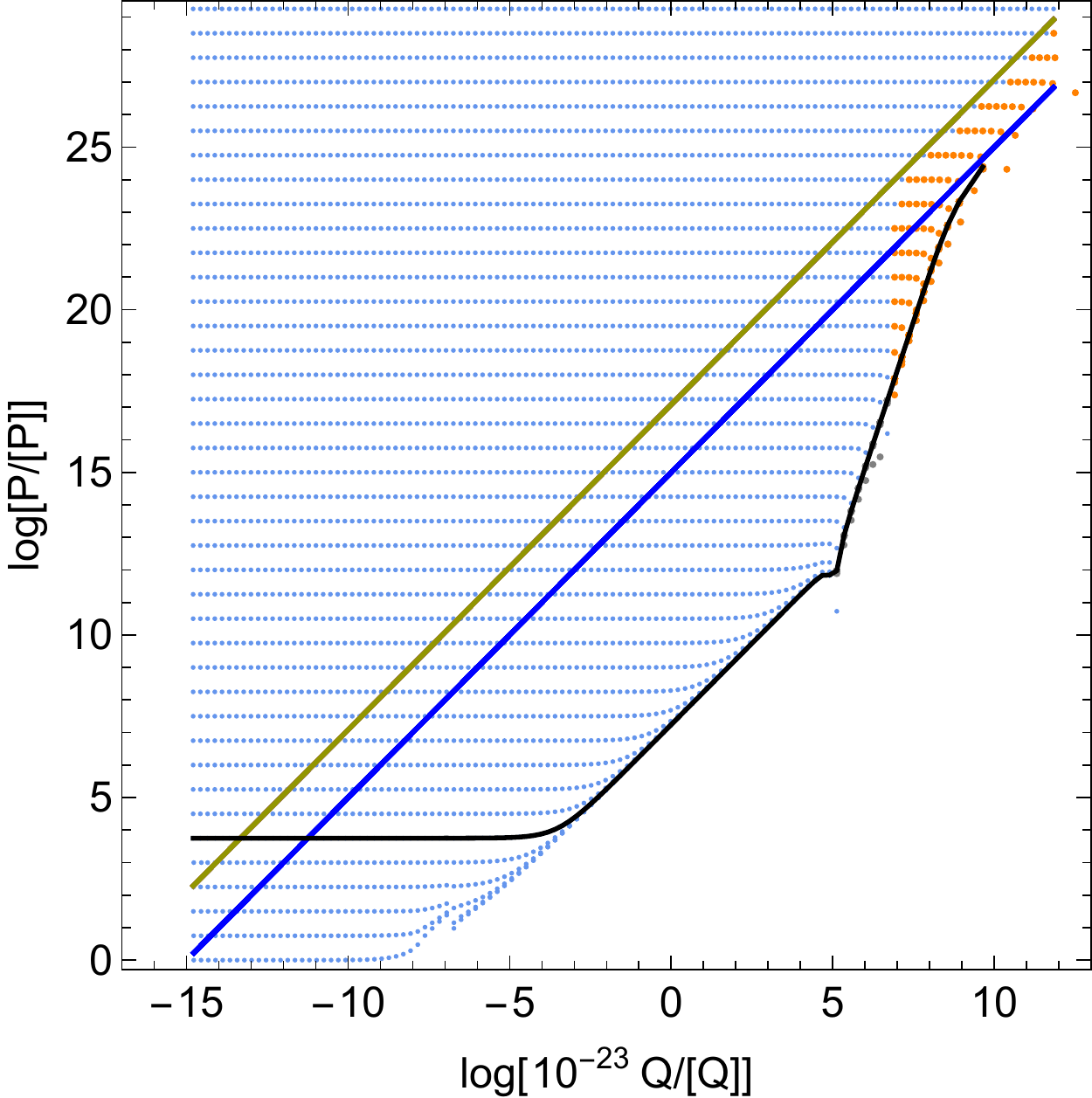}
\caption{ Black hole behaviour in terms of the vector field charges. The axes in the left panel
    are parametrised in terms of quantities evaluated close to the source, $\mc P$ and $\mc Q$, while in the right panel the axes 
    represent the asymptotic quantities $Q$ and $P$. The blue solid line corresponds to $\mc P$ = $\mc Q/r_i$, and the brown solid line to
    $M=P/Q$. The reasons to show these lines are: for solutions below the blue line we naively expect that $\mc Q$ is large
    enough to cause a difference between $\mc P$ and $P$, or, in other words, that an asymptotic expansion of the form $P + Q/r$ is not
    yet valid at $r_i$. As for the brown line, when $P$ and $Q$ have opposite signs -- which is not the case in this plot -- solutions below the line $|M| = |P|/|Q|$ are expected to have a complex $\pi$. This does not happen for this set of solutions, but interestingly enough, for large $P$ and $Q$ the same line signals the transition between regular solutions and solutions without an event horizon. In these plots we fixed
    $\mathcal M = 4.5 M_\odot$. The asymptotic mass $M$ is not fixed, when the vector charges are large the difference between $M$ and $\mathcal M$ is at most $~5\%$, i.e. $1-M/(4.5 M_\odot) \approx 0.05$.
%
%
 }
 \label{f:inics2}
\end{figure}

When we discussed the parameter space for black hole solutions we used the parameters $\mathcal P$ and $\mathcal Q$
defined for numerical convenience as
\begin{align}
\mathcal P &= A_0(r_i) + r_i A_0'(r_i), \\
\mathcal Q &= -\frac{A_0'(r_i)}{r_i^2}.
\end{align}
These parameters correspond to the integration constants $P$ and $Q$ only when they are evaluated at $r_i\to\infty$. Numerically,
we used $r_i = 10^3$ Km. Here we show in what region of the parameter space this $r_i$ is large enough for $\mathcal P$ and $\mathcal Q$ to accurately represent $P$ and $Q$. To do so in a way that is easy to visualise we proceed as follows:
\begin{enumerate}
\item For each solution in Fig.~\ref{f:inicsA} (which is reproduced in the left panel of Fig.~\ref{f:inics2}) we solve from $r_i = 10^3$ Km outwards, up to $r_a = 10^9$ Km, where we can guarantee that the fields have reached their asymptotic values.
\item At $r=r_a$ we compute $P$ and $Q$ for each solution and we redo Fig.~\ref{f:inicsA} parametrising each solution in terms of $P$ and $Q$. This is shown in the right panel of Fig.~\ref{f:inics2}, the colouring in this plot has the same
meaning as in the left panel. The solutions in red in the left panel do not have a corresponding point in the right panel since they do not have an asymptotic region where
$P$ and $Q$ can be defined
\item By comparing the left and right panels of Fig.~\ref{f:inics2} we get a qualitative idea of how different are $(\mathcal P, \mathcal Q)$ and $(P,Q)$.
\end{enumerate}
 For most of the parameter space where regular solutions exist (R4), the ``charges'' evaluated close to the source 
 are a good approximation for the asymptotic values of $P$ and $Q$. Stronger asymptotic effects could be present in 
 the sub-regions of R4 where the blue points in the right panel of Fig.~\ref{f:inics2} are shifted with respect to the left panel,
 a more detailed numerical and analytical study of the solutions in those regions would be required to explore this.

\section{Solutions with  magnetic fields}\label{magneticsolutions}
 
                   In this Appendix we present solutions to the system of vector Galileons \eref{eq:action}
                  in presence of a $\varphi$-component of $A_\mu$ in the branches $\pi = 0$ and $\pi \neq 0$ (which we interpret as `magnetic field' since it turns on the spatial components $F_{ij}$). In both cases,
                  our procedure consists in turning on $A_\varphi$ and identifying the metric components that are necessary to solve
                  the equations of motion at the lowest order in an asymptotic approximation. We  show that \eref{eq:action} admits a richer set of solutions when magnetic configurations are investigated. 
                  
                  \subsection*{Branch $\pi \neq 0$}
                  In this case the component of the metric that we need to turn on to compensate for $A_\varphi$ is $g_{r\varphi}$. 
                  \begin{subequations}
                  \begin{align}
                  1-\frac{2\,n(r)}{r} & = 1 - \epsilon \frac{2 M}{r } \, ,\\
                  1-\frac{2 \, m(r)}{r}  & = 1 - \epsilon \frac{2 M}{r} \, ,\\
                  \hspace{1em}g_{r\varphi}(r,\theta) & =  \epsilon ^{3/2}\frac{2 \sqrt{2} B  \sqrt{M P^2+Q P } \beta  \epsilon ^{3/2} \kappa  \cos\theta}{\sqrt{r} \left(1+P^2 \beta  \kappa \right)}\, , \\
                  \hspace{2em}A_0(r) & = P+\epsilon\frac{Q  }{r}\, ,\\
                  \hspace{2em}\pi(r) & = \epsilon^{1/2}\frac{\sqrt{2}  \sqrt{M P^2+Q P }}{\sqrt{r}}\, , \\
                  \hspace{1em }A_\varphi(r,\theta) & = \epsilon\,  B \cos\theta\, .
                  \end{align}
                  \end{subequations}
                  The metric is still static, but the cross term $G_{r\varphi}$ induces a deformation of the spatial line element that breaks spherical 
                  symmetry, as can be confirmed by computing the invariants $R_{\mu\nu}R^{\mu\nu}$ and $R_{\mu\nu\alpha\beta}R^{\mu\nu\alpha\beta}$
                  and noticing that they depend on $r$ and $\theta$.  Indeed, the Kretschmann scalar is the same as for the Schwarzschild geometry to leading order
                  in $\epsilon$, the angular dependence enters at next-to-leading order and gives a contribution of the 
                  form $w(\theta)/r^7$, where $w(\theta)$ is a function only of $\theta$ that diverges near $\theta = 0$ and $\theta = \pi$, signalling that 
                  near the poles our approximation is only valid for $r\to\infty$.
                  \subsection*{Branch $\pi = 0$}
                  Here we need to turn on the metric components $g_{t\varphi}$ and $g_{r,\theta}$, and the effect of the magnetic field
                  on the other fields is only seen at quadratic and higher orders in $\epsilon$. For simplicity, we set $P=0$, but we checked that
                  an approximated solution can be found with small $P$ as well. For large $P$, the time component of the metric acquires a
                  complicated angular dependence.
                  \begin{subequations}
                  \begin{align}
                  1-\frac{2\,n(r)}{r} & =1-\epsilon\frac{2 M  }{r}+\epsilon ^2 \kappa\frac{B^2+Q^2 (1-4 \beta )   }{2 r^2}\, ,\\
                  1-\frac{2 \, m(r)}{r}  & =1-\epsilon\frac{2 M  }{r}+\epsilon ^2 \kappa\frac{   Q^2+B^2 (1+2 \beta )-2 B^2 \beta  h(\theta)}{2 r^2} \, ,\\
                  \hspace{2.5em}h(\theta) & = \csc^2\theta+\cos\theta  \log\cot\left({\theta }/{2}\right)\, \\
                  \hspace{1em}g_{r\theta}(r,\theta) & =\epsilon ^2 \kappa\, \beta\, B^2   \frac{   \left(11 \cos\theta -3 \cos(3 \theta)+4 \log\tan\left({\theta }/{2}\right) \sin^4\theta\right)}{8 r \sin^3\theta}\, , \\
                  \hspace{1em}g_{t\phi}(r,\theta) & = \epsilon ^2\frac{2\, \kappa\, \beta B\, Q      \cos\theta}{r} \\
                  \hspace{2em}A_0(r) & =\epsilon\frac{Q  }{r}\, ,\\
                  \hspace{1em }A_\varphi(r,\theta) & = \epsilon\,  B \cos\theta\, .
                  \end{align}
                  \end{subequations}
                  
          Once again, the curvature invariants depend on $r$ and $\theta$ and near the poles the approximation is only valid if $r\to\infty$.
          This branch is connected to the magnetic RN solution in the limit $\beta\to 0$.
          
  \section{Currents inside the star: the basic formalism}\label{app:currents}
  The system (\ref{eq:action}) can be generalised in order to explore the consequences of adding a current density term, so that the action now reads
  \begin{equation}
  \label{eq:actionwc}
  S = \int d^4x \sqrt{-g} \left[\frac{1}{2 \kappa}\left(R - 2 \Lambda \right) - \frac{1}{4}F^{\mu\nu}F_{\mu\nu} +  \beta G_{\mu\nu}A^\mu A^\nu  +  j^\mu A_\mu + \mathcal{L}_{matter} \right].
  \end{equation} 
  For simplicity, the current density is chosen to be proportional to the matter density,
  \begin{equation}
  \label{eq:jmu}
  j^\mu =\sqrt{\kappa} \gamma \rho(r) U^\mu ,
  \end{equation}
  where $U^\mu$ is the 4-velocity of the perfect fluid described by $T_{\mu\nu}$.  In the
  rest frame of the perfect fluid -- the same that we use to write the components of
  $T_{\mu\nu}$ -- the only non-vanishing component of the 4-velocity is $U^t =
  \sqrt{g^{tt}}$, where $g^{tt}$ comes from \eref{ansatz}.  $\kappa$ is introduced for
  convenience to make the arbitrary constant $\gamma$ dimensionless: $U^\mu$ is
  dimensionless, $[\rho] = [c^2]\cdot \rm{mass}/\rm{lenght}^3=[\kappa^{-1}][\ell^{-2}]$,
  and $j^\mu$ has to have dimensions of $[A^\mu][\ell^{-2}] =
  [\kappa^{-1/2}][\ell^{-2}] 
  $, these dimensions are given by $\sqrt{\kappa} \rho$.
  The dimensions of $j^\mu$ can be written as $\rm{(electric\ charge)}/\rm{length}^3$,
  therefore the quantity $c^2 j^0/\rho(r) = c^2 \sqrt{\kappa}\gamma U^0$ is a
  mass-to-charge ratio characteristic of the solutions under consideration. Since
  $U^0<<1$, the maximum mass-to-charge ratio is attained at the center of the star and
  given by $c^2 \sqrt{\kappa}\gamma$. 

Let us derive the equations of motion.
          The term $\sqrt{-g} j^\mu$ is constant under variation of the metric. Therefore,
          $j^\mu A_\mu$ only contributes to the vector equation of motion:
          \begin{equation}
          \label{eq:vemc}
          0 = D^\mu F_{\mu\nu} + 2 \beta G_{\mu\nu} A^\mu + j_\nu\, .
          \end{equation}
          Note that, whatever the form of $j^\mu$ -- i.e., forget \eref {eq:jmu} for a
          moment
          , the angular components of this equation demand $j^\theta = j^\varphi = 0$. In
          addition, in the branch $ \pi \neq 0 $ we need to satisfy \eref{eq:mofn} (the
          only assumptions behind this equation, are that $T_{\mu\nu}$ is a perfect fluid
          described in its rest frame and that the metric is diagonal). Once
          \eref{eq:mofn} is substituted in (\ref{eq:vemc}), the $r$-component of such
          equations demands $j^{r}=0$. Thus, $j^0$ is the only new function introduced in
          (\ref{eq:actionwc}).

          Restoring \eref{eq:jmu}, and supplementing our system with the EoS \eref{eosp},
          we have a complete set of equations to determine $A_0(r), n(r)$ and
          $\chi(r)$. The only difference with respect to the system analysed in
          Sec. \ref{sec:ns} is the contribution of $j^0$ to the vector equation.

          Although not necessary --since all the information is encoded in the system of
          equations described above -- it is illuminating to combine the Bianchi identity
          with the vector equation in such a way that the physical difference between the
          systems described by actions \eref{eq:action} and \eref{eq:actionwc} is
          highlighted:
          \begin{equation}
          \label{eq:cont}
          0 =  {j_0} A_0'-{p'} - \frac{n-r\, n'}{r^2-2\, r\,n} (p+ \rho)  \, . 
          \end{equation}
          Eq.~(\ref{eq:cont}) shows that the inclusion of a current density allows the
          vector field to affect directly the equation that determines the matter
          equilibrium configurations. This equation is known as the
          \emph{hydroelectrostatic equilibrium equation}.

          Further physical insight can be gained by rewriting the $t$-component of the
          vector equation of motion as
          \begin{equation}
            0 = j_0(r) - \frac{1}{r^2}\sqrt{-g^{tt}g^{rr}}\left[r^2 \sqrt{-g_{tt}g_{rr}}F^{tr} \right]' - \beta \frac{2 g^{tt} A_0}{r^2} \left[r(1-g^{rr})\right]'\, ,
          \end{equation} 
          where $F^{tr} = -g^{tt}g^{rr}A_0'$. The last expression can be formally
          integrated once to give
          \begin{equation}
            0 = \Theta(r) - r^2\sqrt{-g_{tt} g_{rr}}F^{tr}-\int \frac{2 \beta A_0}{\sqrt{-g_{tt}g^{rr}}}\left[r(1-g^{rr}) \right]' dr,
            \label{eq:theta}
          \end{equation}
          where an integration constant has been absorbed in $\Theta(r) = \int_0^r r^2
          \sqrt{-g_{tt}g_{rr}} j_0 dr + \Theta_0$. Notice that since $j_0\neq 0$ only
          inside a source, $\Theta=\Theta_0$ for exterior solutions. The last term in
          \eref{eq:theta} is in general non-vanishing, $A_0$ sources the electromagnetic
          tensor and consequently the gravitational fields in and outside the star. Far
          from the source, in asymptotically flat spacetime, (\ref{eq:theta}) gives as
          solution $A_0 = \Theta_0/r + const.$, thus $\theta_0$ can be identified with the
          asymptotic charge $Q$. The asymptotic corrections discussed around
          eq. (\ref{a0app}) originate from the last term in \eref{eq:theta}.
\end{appendix}


\end{document}